\documentclass[fleqn,usenatbib]{mnras}

\usepackage{newtxtext,newtxmath}

\usepackage[T1]{fontenc}
\DeclareRobustCommand{\VAN}[3]{#2}
\let\VANthebibliography\thebibliography
\def\thebibliography{\DeclareRobustCommand{\VAN}[3]{##3}\VANthebibliography}

\usepackage{float}
\usepackage{caption}
\usepackage{subcaption}

\usepackage{graphicx}	
\usepackage{amsmath}	

\title[\textsc{fried} v2]{\textsc{fried} v2. A new grid of mass loss rates for externally irradiated protoplanetary discs}

\author[T. J. Haworth et al. ]{
Thomas J. Haworth,$^{1}$\thanks{E-mail: t.haworth@qmul.ac.uk},  Gavin A. L. Coleman$^1$, Lin Qiao$^1$,  Andrew D. Sellek$^2$  and Kanaar Askari$^1$
\\
$^{1}$Astronomy Unit, School of Physics and Astronomy, Queen Mary University of London, London E1 4NS, UK\\
$^{2}$Institute of Astronomy, University of Cambridge, Madingley Road, Cambridge CB3 0HA, UK\
}

\date{Accepted XXX. Received YYY; in original form ZZZ}

\pubyear{2023}

\begin{document}
\label{firstpage}
\pagerange{\pageref{firstpage}--\pageref{lastpage}}
\maketitle

\begin{abstract}
We present a new \textsc{fried} grid of mass loss rates for externally far-ultraviolet (FUV) irradiated protoplanetary discs. As a precursor to the new grid, we also explore the microphysics of external photoevaporation, determining the impact of polycyclic aromatic hydrocarbon (PAH) abundance, metallicity, coolant depletion (via freeze out and radial drift) and grain growth (depletion of small dust in the outer disc) on disc mass loss rates. We find that metallicity variations typically have a small effect on the mass loss rate, since the impact of changes in heating, cooling and optical depth to the disc approximately cancel out. The new \textsc{fried} grid therefore focuses on i) expanding the basic physical parameter space (disc mass, radius, UV field, stellar mass) ii) on enabling variation of the the PAH abundance and iii) including an option for grain growth to have occurred or not in the disc. What we suggest is the fiducial model is comparable to the original \textsc{fried} grid. When the PAH-to-dust ratio is lower, or the dust in the wind more abundant, the mass loss rate can be substantially lower. We demonstrate with a small set of illustrative disc evolutionary calculations that this in turn can have a significant impact on the disc mass/radius/ evolution and lifetime. 
\end{abstract}




\section{Introduction}
Planet-forming discs are found around young stars \citep{2007ApJ...671.1784H, 2010A&A...510A..72F, 2015A&A...576A..52R} which in turn generally reside in stellar clusters \citep[e.g.][]{2003ARA&A..41...57L, 2008ApJ...675.1361F, 2020MNRAS.491..903W}. In the larger clusters where most stars are formed, massive stars emit large amounts of UV radiation which can disperse the star forming cloud \citep[e.g.][]{2006ApJ...647..397M, 2013MNRAS.430..234D, 2013MNRAS.435..917W, 2015NewAR..68....1D, 2020MNRAS.495.1672B, 2021MNRAS.501.4136A, 2021MNRAS.506.2199G, 2022MNRAS.509..954D}.  This UV radiation can also strip material from protoplanetary discs, which is referred to as external photoevaporation (with the term ``external'' distinguishing it from photoevaporation by the disc's host star, which we call ``internal'' photoevaporation). For recent reviews on internal and external photoevaporation see \cite{2022arXiv220310068P} and \cite{2022EPJP..137.1132W} respectively.

Observationally there is evidence supporting the idea that external photoevaporation can have a significant effect on the evolution of planet-forming discs. The least subtle evidence is that in star forming regions in Orion ($\sim400$\,pc distant) the externally photoevaporating discs take on a cometary morphology due to the winds being driven from them, and we refer to these objects as ``proplyds'' \citep{1993ApJ...410..696O, 1994ApJ...436..194O, 1998AJ....115..263O, 1999AJ....118.2350H, 2000AJ....119.2919B, 2001AJ....122.2662O, 2008AJ....136.2136R, 2016ApJ...826L..15K, 2021MNRAS.501.3502H, 2022EPJP..137.1132W}. Combining the incident ionising flux and the radius of the ionisation front, one can estimate the mass loss rate from proplyds. These are routinely inferred to be $\sim10^{-6}$\,M$_\odot$\,yr$^{-1}$, which implies disc depletion timescales ($M/\dot{M}$) of order 0.1\,Myr \citep[e.g.][]{1999AJ....118.2350H}. This is a lower limit on the true remaining disc lifetime, as the mass loss rate is a function of disc radius and the external far ultraviolet (FUV) radiation field strength, which both change with time \citep{2007MNRAS.376.1350C, 2013ApJ...774....9A,2019MNRAS.490.5678C, 2022MNRAS.512.3788Q, 2023MNRAS.520.5331W} but regardless it illustrates that the inferred mass loss rates are at a level that is significant. 

In addition to direct evidence of externally photoevaporating discs, there is also statistical evidence that the properties of discs {vary as a function of the FUV field strength}\footnote{{The far ultraviolet (FUV) radiation field strength is usually measured using the Habing unit,  defined as $1$G$_0=1.6\times10^{-3}$ erg\,s$^{-1}$\,cm$^{-2}$ integrated over the wavelength range 912--2400\AA. 1\,G$_0$ is representative of the mean interstellar FUV radiation field in the Solar neighbourhood. }}. There is evidence for shortened disc lifetimes \citep{2016arXiv160501773G, 2020A&A...640A..27V}, smaller disc radii \citep{2018ApJ...860...77E, 2020ApJ...894...74B, 2023ApJ...947....7B}, lower disc masses \citep{2017AJ....153..240A}, and even different gas composition \citep{2023ApJ...947....7B} in higher UV environments. There is also recent evidence for {gradients in disc masses in intermediate--to--low}\footnote{{As a rough guide we refer to low, intermediate and high FUV environments as F$_{\textrm{FUV}}<10^2$G$_0$, $10^2$G$_0 < $ F$_{\textrm{FUV}} < 10^4$G$_0$ and F$_{\textrm{FUV}} > 10^4$\,G$_0$ respectively.}} UV environments where there is no observed cometary proplyd morphology \citep{2023arXiv230405777V}. The interpretation of these statistical trends is not without ambiguity, but coupled with direct disc observations the evidence for external photoevaporation being important for disc evolution is continuing to grow.   

Unfortunately, computing the rate at which material is lost from an externally irradiated disc is difficult because the temperature of the irradiated gas requires solving for photodissociation region (PDR) chemistry \citep{2004ApJ...611..360A, 2016MNRAS.457.3593F, 2016MNRAS.463.3616H}. This is rather computationally expensive, particularly if doing 2D or 3D simulations which require computing the degree of line cooling in many different directions from each point in the simulation. For internally driven winds or inside the disc it is usually assumed that the escape of line photons is mainly perpendicular to the disc \citep[e.g.][]{2008ApJ...683..287G, 2019ApJ...874...90W} but that doesn't necessarily apply in an external wind. It is therefore currently impossible to simulate a disc for an appreciable fraction of its lifetime (i.e. for millions of years) and directly work out how it is being externally photoevaporated simultaneously. What we can do though is pre-compute the mass loss rate for a large number of star-disc-UV field combinations and then interpolate over those in a much faster disc evolutionary model. This allows us to evolve discs for millions of years, including the effect of external photoevaporation \citep[e.g.][]{2013ApJ...774....9A, 2017MNRAS.468L.108H, 2018MNRAS.475.5460H}. 

To address the above issue and open up the modelling of externally photoevaporating discs to the community, \cite{2018MNRAS.481..452H} produced the \textsc{fried} ({F}ar-ultraviolet {R}adiation {I}nduced {E}vaporation of {D}iscs) grid of mass loss rates. For an input stellar mass, disc radius, disc mass (or surface density) and incident UV field, \textsc{fried}  returns an external photoevaporative mass loss rate. \textsc{fried} has been widely used, particularly for studying the evolution of discs in different environments \citep{2018MNRAS.478.2700W, 2020MNRAS.491..903W, 2020MNRAS.492.1279S, 2020MNRAS.496L.111C, 2022MNRAS.509...44W, 2022MNRAS.514.2315C, 2022ApJ...926L..23H} including time evolving UV fields due to cluster and interstellar medium dynamics \citep{2019MNRAS.485.4893N, 2019MNRAS.490.5478W, 2019MNRAS.490.5678C, 2021MNRAS.501.1782C, 2021ApJ...913...95P,2021MNRAS.502.2665P, 2022MNRAS.512.3788Q, 2022MNRAS.515.5449M, 2022MNRAS.517.2103D, 2023MNRAS.520.5331W, 2023MNRAS.520.6159C}.  It has also recently been used to demonstrate that external photoevaporation could influence the properties of giant planets \citep{2022MNRAS.515.4287W} and planets formed by pebble accretion \citep{2023arXiv230315177Q}, as well as for studying planet formation in circumbinary discs \citep{2023MNRAS.tmp..787C}. It is also featuring now in planet population synthesis codes \citep{2021A&A...656A..72B, 2022A&A...666A..73B, 2023arXiv230104656E, 2023EPJP..138..181E}. \textsc{fried} was also used to demonstrate that shortened disc lifetime in high UV environments could affect stellar rotation rates \citep{2021MNRAS.508.3710R}. 

This paper presents a second version of the \textsc{fried} grid, which builds on the original by i) covering a wider parameter space than the first, ii) providing mass loss rates down to lower values than the previous floor value, iii) allowing one to account for whether grain growth in the disc has proceeded to the disc outer edge and iv) allowing one to choose the polycyclic aromatic hydrocarbon (PAH) abundance, which can be the dominant heating mechanism in the PDR. In addition to detailing and presenting the new grid, section \ref{sec:lessons} of this paper provides a guide to utilising \textsc{fried} in disc evolutionary models, informed by the lessons learned in applications to date. In section \ref{sec:discussion} we also present some initial illustrative disc evolutionary calculations.

\section{1D models of external photoevaporation with \textsc{torus-3dpdr}}
The calculations in this paper use the \textsc{torus-3dpdr} code \citep{2015MNRAS.454.2828B, 2019A&C....27...63H}. Much of the approach is the same as in other applications of this code to disc photoevaporation in recent years, including the original \textsc{fried} grid \citep{2018MNRAS.481..452H}. However there are also some new additions in order to study and interpret the impact of microphysics. We therefore provide only a brief overview of the core methodology and focus on new developments here.

The external photoevaporation calculations with \textsc{torus-3dpdr} run here are 1D grid based models of hydrodynamics with a locally isothermal equation of state where the temperature is set by photodissociation region (PDR) physics. The grid itself uses adaptive mesh refinement \citep[AMR,][]{2019A&C....27...63H}.  The PDR and hydrodynamic steps are done iteratively (i.e. using operator splitting). The PDR calculation uses a reduced UMIST network \citep{2013A&A...550A..36M} with 33 species and 330 reactions, tailored to give temperatures within $\sim10$\,per cent of the full network (215 species $>3000$ reactions). {For further information on the network and heating and cooling processes see \cite{2012MNRAS.427.2100B} and we note that the heating and cooling processes are the same as summarised in Figure 2 of \cite{2016MNRAS.457.3593F}}. 

Radiation in the 1D PDR calculation is assumed to be purely radial, with UV radiation incident on the disc outer edge and cooling radiation escaping in the opposite direction. All other lines of sight are assumed to be infinitely optically thick. This 1D approach is the same as used by e.g. \cite{2004ApJ...611..360A} and \cite{2016MNRAS.457.3593F} and was validated in comparison to 2D models (at least for computing mass loss rates) by \cite{2019MNRAS.485.3895H}, who demonstrated that 1D models give slightly lower mass loss rates than the 2D calculations. 

The hydrodynamics is total variation diminishing finite volume scheme with \cite{1983AIAAJ..21.1525R} interpolation and a \cite{vanleer} flux limiter. Given the simple geometry of these models we just consider a point source potential set by the host star, so do not treat self-gravity. 

In these calculations a protoplanetary disc acts as a fixed inner boundary condition that is not allowed to evolve. We define this inner boundary based on the surface density normalization at 1\,au $\Sigma_{1\textrm{au}}$ and the disc radius assuming a surface density profile of the form
\begin{equation}
    \Sigma(R) = \Sigma_{1\textrm{au}}\left(\frac{R}{\textrm{au}}\right)^{-1}
\end{equation}
and temperature profile as set by the disc host star of the form 
\begin{equation}
    T(R) = T_{1\textrm{au}}\left(\frac{R}{\textrm{au}}\right)^{-1/2}
\end{equation}
{with a floor value of 10\,K.}
The mid-plane density profile modelled on the 1D grid is then set by $\rho_{\textrm{mid}} = \frac{\Sigma}{\sqrt{2\pi}H}$ where $H=c_s/\Omega$ is the disc scale height (the ratio of sound speed to angular velocity). $T_{1\textrm{au}}$ is assumed to scale with the host star mass $M_*$ as 
\begin{equation}
    T_{1\textrm{au}} = 100\left(M_*/M_\odot  \right)^{1/4}
\end{equation}
{which stems from a simple $L\propto M_*$ pre main sequence mass-luminosity relationship and radiative equilibrium.  Stellar pre main sequence evolution and accretion may alter the temperature structure of the passively heated disc, but this is time variable and adds complexity to the model beyond the scope of this paper that would not significantly influence the results.}
Beyond the fixed inner boundary described above the medium is allowed to evolve towards a steady state wind solution. 

This approach applies best in the regime where the disc outer edge is optically thick to to the incident FUV radiation. It is possible that our calculations can set up a scenario where the disc outer edge is optically thin, which is artificial because in reality the FUV would penetrate deeper into the disc and drive the wind from smaller radii.  A slim-disc approach to this problem was introduced by \cite{2021MNRAS.508.2493O} which circumvents that issue and naturally connects the disc and an isothermal wind. We retain our approach and note that in the regime that the outer disc is optically thin it evolves rapidly into the optically thick regime ($\Sigma(R_d)/\dot{\Sigma}$ is short).

The disc boundary condition is irradiated and the PDR-dynamics run until a steady state wind solution is reached. The mass loss in the wind is calculated from the 1D model by assuming a spherical wind into the solid angle subtended by the disc outer edge \citep{2004ApJ...611..360A}. That is, the mass loss rate at some point $R$ in the flow of density $\rho$ is 
\begin{equation}
    \dot{M} = 4\pi R^2 \dot{R} \rho \mathcal{F}
    \label{equn:Mdot}
\end{equation}
where
\begin{equation}
    \mathcal{F} = \frac{H_d}{\sqrt{H_d^2+R_d^2}}
    \label{equn:mathcalF}
\end{equation}
is the solid angle subtended by the disc outer scale height $H_d$ at radius $R_d$. The scale height at the disc outer edge is dependent on the sound speed (and hence temperature) there, which is the maximum of the temperature set by the host star or the external irradiation. 

The species included in our models is summarised in Table \ref{table:speciesparams}. The cosmic ray ionisation rate is assumed to be $5\times10^{-17}$\,s$^{-1}$. The other microphysical parameters to do with the dust and PAH abundances are discussed below. In all calculations the grid extent is 2000\,au with a maximum spatial resolution with the AMR of 0.97\,au. 

\begin{table}
 \centering
  \caption{The 33 species included in the reduced network of our PDR-dynamical models \protect\citep{2009ARA&A..47..481A}. The sum of  hydrogen atoms in atomic and molecular hydrogen is unity. The other abundances are with respect to the sum of hydrogen parameters. }
  \label{table:speciesparams}
  \begin{tabular}{@{}l c l c@{}}
   \hline
   Species & Initial abundance & Species & Initial abundance \\
   \hline
   H & $4\times10^{-1}$ & H$_2$ & $3\times10^{-1}$ \\
   He & $8.5\times10^{-2}$ & C+ & $2.692\times10^{-4}$\\
   O & $4.898\times10^{-4}$ & Mg+ & $3.981\times10^{-5}$ \\
   H+ & 0 & H$_2$+ & 0 \\
   H$_3$+ & 0& He+ & 0 \\
   O+ & 0 & O$_2$ & 0 \\
   O$_2$+ & 0 & OH+ & 0 \\
   C & 0 & CO & 0 \\
   CO+ & 0 & OH & 0 \\
   HCO+& 0 & Mg & 0 \\
   H$_2$O& 0 & H$_2$O+ & 0 \\
   H$_3$O & 0 & CH & 0 \\
   CH+ & 0 & CH$_2$ & 0 \\
   CH$_2$+ & 0 & CH$_3$ & 0 \\
   CH$_3$+ & 0 &CH$_4$ & 0 \\ 
   CH$_4$+ & 0 & CH$_5$+ & 0 \\    
   e$^-$  & 0 \\      
\hline
\end{tabular}
\end{table}

\section{The microphysics of external photoevaporation}
\label{sec:microphysics}
There are many variables that enter into the calculations of external photoevaporative mass loss. Some are macroscopic, such as the star-disc-UV parameters, which were explored in the first \textsc{fried} grid \citep{2018MNRAS.481..452H}. However other important parameters are to do with the microphysics such as the metallicity, degree of dust grain growth/entrainment in the wind and the PAH abundance. We therefore undertook a preliminary study with a smaller subset of models in which we varied a number of microphysics parameters to determine their expected importance, and hence decide what to prioritise in the new \textsc{fried} grid. Note though, that this exploration is limited in scope and so does not exhaustively demonstrate the relative importance of different parameters. 

{Before we go into the parameter exploration, we first quickly recap what are generally the key microphysics processes in these models. The included and relevant importance of heating and cooling processes is summarised in Figure 2 of \cite{2016MNRAS.457.3593F}.  In a fiducial model the key heating process is photoelectric heating by PAH's, except for at high extinction where cosmic rays, turbulence and gas-grain heating takes over. Other heating processes include carbon ionisation, H$_2$ formation and photodissociation, FUV pumping and other chemical heating processes. {Line cooling is included from CO, C, C$^+$ and O. CO dominates the cooling in the very inner subsonic part of the flow where CO is not yet dissociated \citep{2016MNRAS.457.3593F, 2016MNRAS.463.3616H}}, then there is a small layer where C and C$^+$ are important, but the bulk of the flow outside of the CO cooling zone is dominated by O line cooling.  The other key factor, which is also discussed in \cite{2016MNRAS.457.3593F}, is the extinction (and hence properties of the dust) in the wind. This affects the attenuation of the incident UV radiation and hence the ability of the external radiation field to heat the outer disc.  }

\subsection{Metallicity and PAH-to-dust mass ratio}
\label{sec:metalScale}
One of the most obvious features missing from \textsc{fried} is the option to vary the metallicity (e.g. the abundance of gas metals, dust and PAH's in a consistent manner). However, even for fixed metallicity the abundance of the species responsible for the (potentially) dominant heating channel -- photoelectric heating by PAH's -- is poorly constrained in the outer region of discs. PAH's are molecules dominated by carbon rings with delocalised electrons that are quite easily ejected, heating the gas \citep[see][for a review]{2008ARA&A..46..289T}. 

We discuss observational constrains on the PAH abundance in proplyds and the implications for using \textsc{fried} in section \ref{sec:fiducial}. However for now we note that the small number of estimates available find that the PAH-to-gas ratio is lower than that in the ISM \citep[though this could also be because the dust-to-gas ratio is depleted in the wind rather than pure PAH depletion,][]{2016MNRAS.457.3593F}. Additionally PAHs could be ``hidden'' in aggregate clusters \citep{2021A&A...653A..21L}.  Ultimately the PAH abudance is uncertain, and so we therefore need to ask whether it is more prudent with finite resources to provide an option to modify the overall metallicity (abundance of gas metals, dust and PAH abundance simultaneously) or rather just modify the PAH abundance.

In \textsc{torus-3dpdr}, scaling of the PAH abundance is implemented in practice as a scaling of the PAH-to-dust mass ratio. The fiducial value of PAH-to-dust mass ratio is $\delta_{\rm PAH}=2.6\times10^{-2}$ following \cite{2003ApJ...587..278W}. We use the parameter $f_{\textrm{PAH,d}}$ to deplete the PAH-to-dust abundance relative to that fiducial value, so $f_{\textrm{PAH,d}}=1$ is an ISM-like \textit{PAH-to-dust} ratio. Observational constraints on the PAH abundance are usually in terms of the \textit{PAH-to-gas} mass ratio  \citep[e.g.][]{2013ApJ...765L..38V}. In this paper we write the PAH-to-gas ratio relative to that in the ISM using $f_{\textrm{PAH,g}}$, so a factor 50 depletion of the PAH-to-gas abundance relative to the ISM would be $f_{\textrm{PAH,g}}=0.02$, but if that were due to a factor 50 reduction in the dust-to-gas mass ratio then it would still be the case that $f_{\textrm{PAH,d}}=1$.  When we reduce both the PAH abundance (i.e. $f_{\textrm{PAH,d}}$) and the metallicity, the PAH abundance is first be scaled down by the factor $f_{\textrm{PAH,d}}$ and then further scaled linearly with the metallicity.

\begin{figure}
    \centering
    \includegraphics[width=\columnwidth]{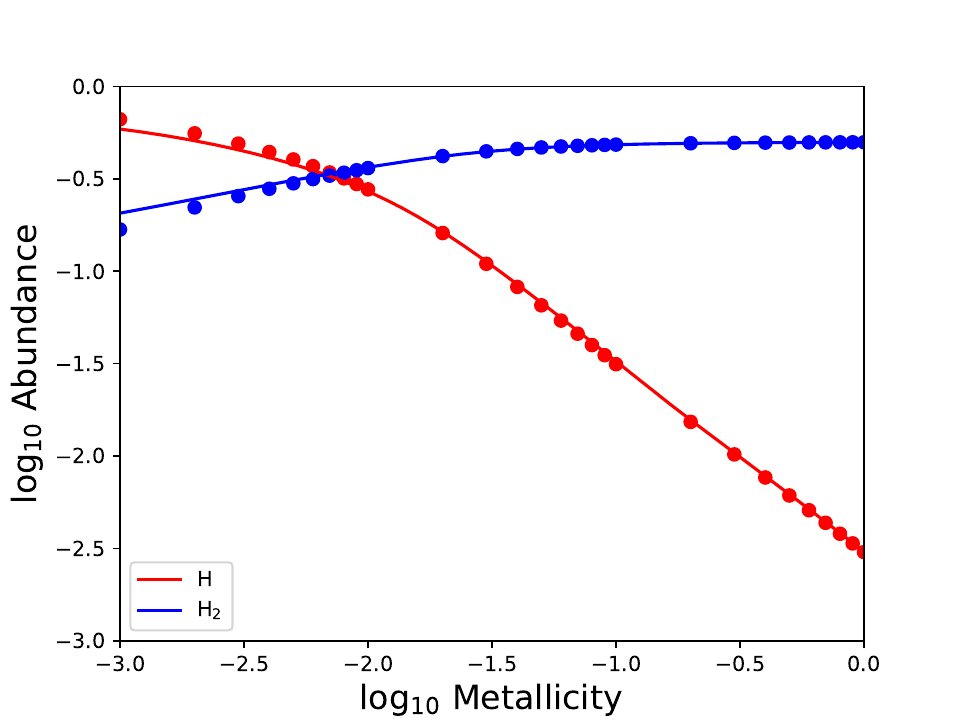}
    \vspace{-0.3cm}
    
    \includegraphics[width=\columnwidth]{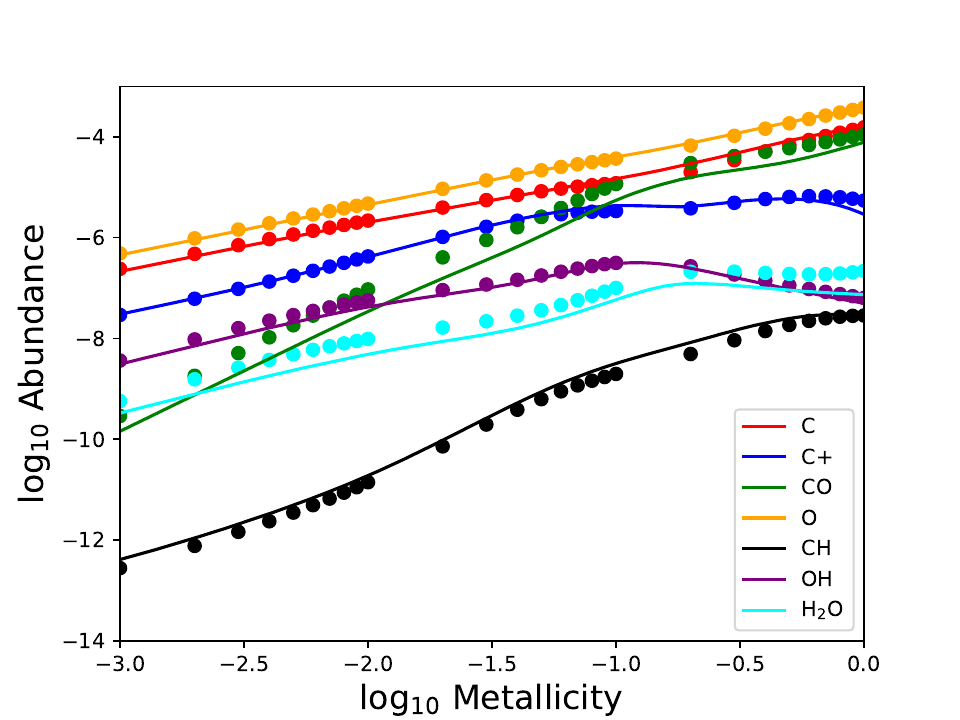}    \caption{Comparison of abundances as a function of metallicity, comparing \textsc{torus-3dpdr} with the  $I_{UV,0}/n_3=1$ model from \protect\cite{2015MNRAS.450.4424B}. Lines are from \protect\cite{2015MNRAS.450.4424B} and points are from \textsc{torus-3pdr} }
    \label{fig:metallicityBenchmark}
\end{figure}

Before exploring the impact on external photoevaporative mass loss rates we {perform a benchmarking calculation of abundances as a function of metallicity to compare our} \textsc{torus-3pdr} calculations against the $I_{UV,0}/n_3=1$ (Draine FUV field/number density in units of $10^3$\,cm$^{-3}$) results of \cite{2015MNRAS.450.4424B}, finding excellent agreement as illustrated in Figure \ref{fig:metallicityBenchmark}.

\begin{figure}
    \centering
    \includegraphics[width=1.1\columnwidth]{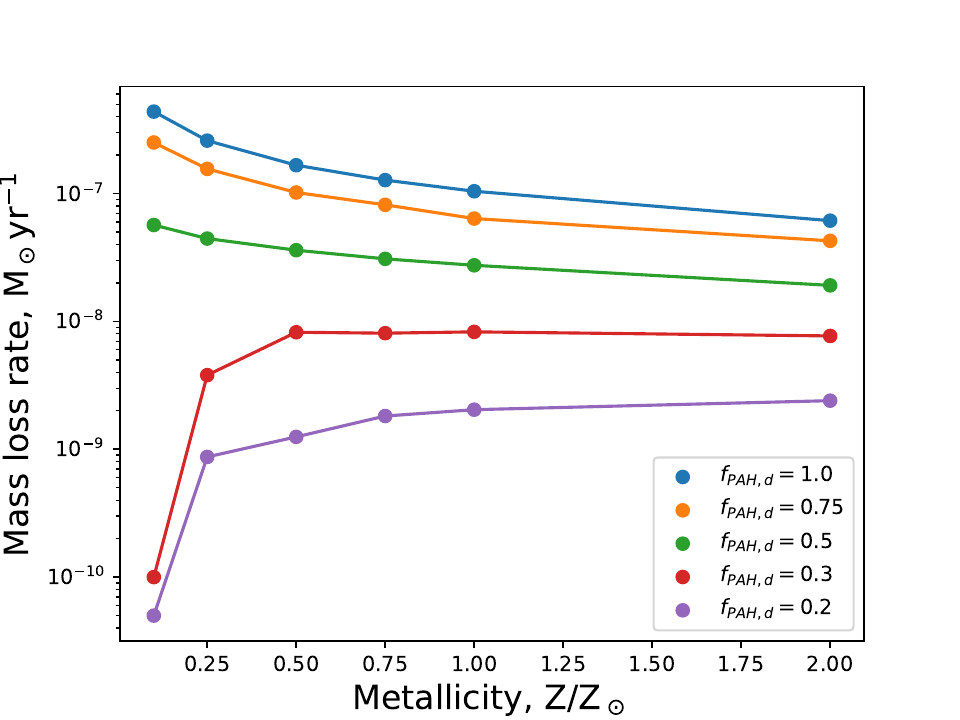}
    \caption{Exploratory calculations of the external photoevaporative mass loss rate as a function of metallicity for a 100\,au disc around a 1\,M$_\odot$ star irradiated by a 10$^3$\,G$_0$ radiation field. Each coloured line represents a different value of the base PAH-to-dust mass ratio scaling.}
    \label{fig:metallicityAndPAH}
\end{figure}

We ran a small set of models comprising a 1\,M$_\odot$ host star with $\Sigma_{1\textrm{au}}=10^3$\,g\,cm$^{-2}$, a $R_d=100$\,au disc and an external FUV radiation field of $10^3$\,G$_0$ at a range of metallicities and values of the base PAH-to-dust mass ratio $f_{\textrm{PAH,d}}$. The results of this are presented in Figure \ref{fig:metallicityAndPAH}, which shows the mass loss rate as a function of metallicity for different base PAH-to-dust mass ratios. The variation of mass loss rate as a function of metallicity is actually rather modest. This is due to an approximate balance between the competing processes that promote and hinder mass loss. {On the one hand, at lower metallicities there is weaker line cooling, less dust in the wind and hence less extinction, which would both act to promote heating and mass loss. However this is offset by the reduction in PAH abundance, which is the most significant process heating the PDR.  Conversely, changing the base PAH-to-dust mass ratio solely changes the heating, without affecting the cooling processes, and hence has a much larger impact on the mass loss rate}. Note that at very low metallicity and low PAH abundance the outer disc is not sufficiently heated to unbind material, which is why the mass loss rate is so much lower. We reiterate that this is just one subset of parameters and sensitivity to both metallicity and PAH abundance will differ in strength for different star/disc/UV field parameters. However, based on this initial exploration it seems that the more prudent parameter to vary is the PAH-to-dust mass ratio rather than the metallicity. 

\subsection{Dust grain entrainment and dust evolution}
\cite{2016MNRAS.457.3593F} determined the maximum size of dust grains that can be entrained in an external photoevaporative wind. Given that this is typically small (e.g. $<10$\,$\mu$m) the implication is that the nature of dust entrained in the wind is dependent upon the degree of grain growth and radial drift \citep{1977MNRAS.180...57W} within the disc itself. If the dust in the disc has grown to larger sizes then the small grain reservoir in the disc is depleted, less dust is entrained in the wind and the disc becomes more optically thin to the FUV radiation. Conversely if the dust in the disc has not undergone growth and drift, being like that in the interstellar medium (ISM-like), then much more dust is entrained in the photoevaporative wind and the attenuation of the FUV radiation is stronger. Although grain growth could proceed very quickly \citep[e.g.][]{2012A&A...539A.148B}, it does so from the inside-out. The early external photoevaporation of discs may therefore take place somewhat less effectively than expected from models that assume grain growth has occurred (i.e. as the original \textsc{fried} grid did) which could be important for helping to protect the solids reservoir for planet formation \citep{2023arXiv230315177Q}.

\begin{figure}
    \centering
    \includegraphics[width=1.1\columnwidth]{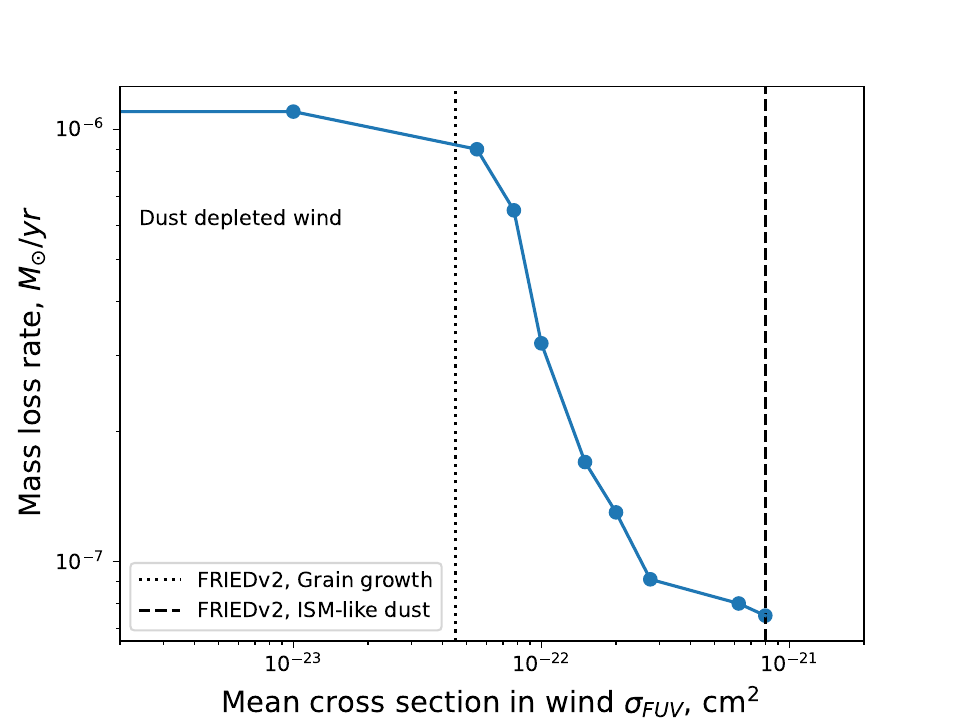}
    \caption{The mass loss rate in an external photoevaporation calculation as a function of the mean cross section in the wind. When dust is ISM-like, more dust is entrained and the mean cross section is larger. When dust growth and radial drift depletes small dust near the disc outer edge the mean cross section in the wind and hence the extinction to the disc outer edge decreases, meaning the mass loss rate increases. The vertical lines show the two choices of mean cross section that we include in the new \textsc{fried} grid. }
    \label{fig:Mdot_crossec}
\end{figure}

Changing the dust entrained in the wind affects the calculation in a few main ways. Firstly it affects the grain surface chemistry in the wind by changing the dust abundance. Secondly, and more significantly, it affects the PAH abundance in the wind (since it is defined as a PAH-to-dust mass ratio in our models). Finally, it changes the effective mean cross section in the wind, which alters the extinction in the flow. 

In Figure \ref{fig:Mdot_crossec} we show the sensitivity of the mass loss rate to that mean cross section to the UV. When the dust in the disc has grown to large sizes, there is less small dust to entrain in the wind, leaving the wind dust depleted (this is the case on the left hand side of Figure \ref{fig:Mdot_crossec}) and the mass loss rate is higher as FUV radiation more effectively reaches down to the disc outer edge. This higher mass loss rate is \textit{in spite} of the reduced PAH abundance due to the reduction in the dust abundance. Conversely, when the dust in the disc is ISM-like the mean cross section in the wind is higher, less FUV reaches the disc outer edge and the mass loss rate drops (by around an order of magnitude in this case), this again is despite the higher PAH abundance in this model. 

Given the importance of the grain properties in the wind we choose two dust scenarios for the new \textsc{fried} grid. In one the dust entrained in the wind is ISM like (i.e. there is little/no grain growth in the disc), and in the other the dust entrained in the wind is depleted (i.e. small grains in the disc have grown to larger sizes that are not entrained). {We refer to these as ``ISM-like'' and ``grain growth'' cases respectively.} The corresponding mean cross sections are illustrated by the vertical dotted and dashed lines in Figure \ref{fig:Mdot_crossec}. To account for less effective external photoevaporation early on in the disc lifetime, a viscous disc model with a    \cite{2012A&A...539A.148B} (or similar) grain evolution scheme could evolve the disc under the ISM-like subset of the new \textsc{fried} grid until the pebble production front has reached the disc outer edge, and then transition to the dust depleted external photoevaporation scheme.

\subsection{Carbon depletion}
{Carbon depletion in the outer disc could be facilitated by freeze out onto grains \citep[e.g][]{1981PThPS..70...35H, 2011ApJ...743L..16O, 2019MNRAS.487.3998B} coupled with radial drift \citep{1977MNRAS.180...57W} moving the carbon to the inner disc so that when the outer disc is externally irradiated it isn't simply desorbed. Depletion of carbon in the disc has already been demonstrated to be significant for internal photoevaporative mass loss rates \citep[since carbon is an opacity source to the X-rays;][]{2019MNRAS.490.5596W}. In the colder outer disc, from which grains drift inwards without replenishment, it would seem only more likely that depletion occurs. It has also been suggested as a possibility in the NGC 1977 proplyds \citep{2022MNRAS.512.2594H}.}

\begin{figure}
    \centering
    \includegraphics[width=1.1\columnwidth]{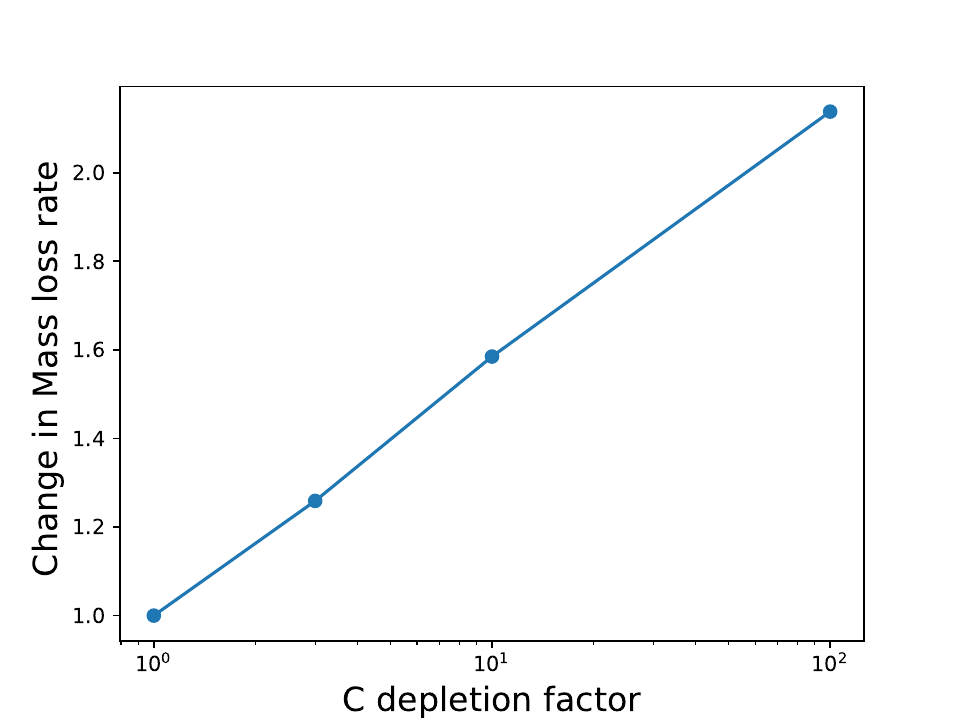}
    \caption{The change in external photoevaporative mass loss rate as the carbon abundance {(and hence cooling from CO, C and C$^+$ which is dominant in the inner wind, Facchini et al. 2016) is depleted. The main coolant throughout the bulk of the wind is O line emission.}}
    \label{fig:coolantMdots}
\end{figure}

{To explore this effect we set up an initial calculation with abundances as in Table \ref{table:speciesparams} and then scaled down the abundance of carbon by some arbitrary factor. The effect of this on the mass loss rate is illustrated in Figure \ref{fig:coolantMdots}, which shows that with 2 orders of magnitude variation in the carbon abundance the mass loss rate changes by only around a factor of 2. As discussed above, the main coolant in the bulk of the wind is atomic oxygen line cooling. So the reduction in carbon abundance only affects the cooling in the CO zone and thin C/C$^+$ cooling layers, which has a small impact on the mass loss rate. We therefore deem that carbon depletion is not as significant a factor to include as changes to the PAH abundance or dust properties in the wind, at least as far as the mass loss rate is concerned. Although we do not prioritise it for calculating mass loss rates, depletion in the outer disc will be very important for understanding observables \citep[][Sellek et al. in prep]{2022MNRAS.512.2594H}. Given that oxygen is the dominant cooling mechanism in the outer wind the sensitivity of mass loss rate to the oxygen abundance should be explored in future work. }

\subsection{Including the EUV}
\textsc{fried} v1 only included photodissociating FUV radiation. It does not include photoionising extreme ultraviolet (EUV) radiation, which is what gives rise to the cometary ionisation-front morphology of proplyds. However \textsc{torus} includes a Monte Carlo photoionisation scheme \citep{2012MNRAS.420..562H} so the EUV can be included quite easily. To include both FUV and EUV radiation we first do a Monte Carlo photoionisation calculation that calculates the photoionised gas temperature, and since it is polychromatic simultaneously calculates the FUV radiation field distribution \citep[e.g. as done by][]{2019MNRAS.487.4890A}. We then do the PDR phase of the calculation, applying the resulting PDR temperature in any cells that are hotter than the photoionisation computed temperature. In these calculations we do not arbitrarily specify a particular external EUV and FUV flux, but calculate both consistently based on a blackbody stellar source at some distance assuming geometric dilution of the radiation field. 

\begin{figure}
    \centering
    \includegraphics[width=1.1\columnwidth]{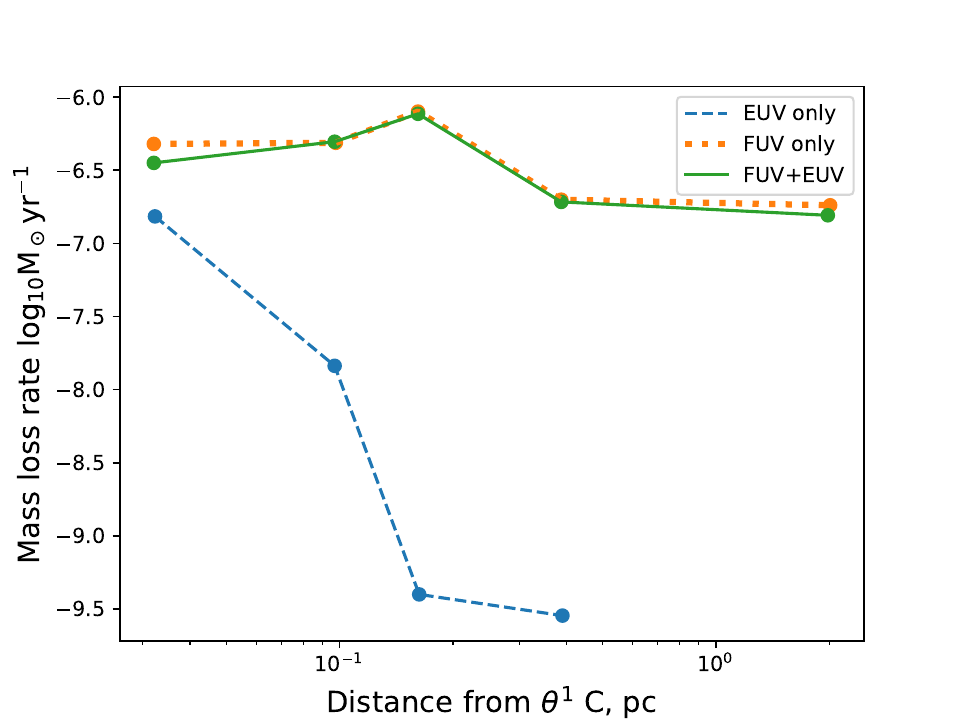}
    \caption{The external photoevaporative mass loss rate as a function of distance of a disc from a $\theta^1$ Ori C like UV source. The FUV is dominant in determining the mass loss rate.   }
    \label{fig:EUVvsFUV}
\end{figure}

For our exploration of the impact of the EUV we consider a massive star that is a blackbody with radius 8\,R$_\odot$ and effective temperature $T_{\textrm{eff}}$=40,000\,K, so similar to the Orion Nebular Cluster Trapezium star $\theta^1$ Ori C. We place a 100\,au disc around a 1\,M$_\odot$ star and move it progressively closer to the massive star, comparing the mass loss rate if we only include the EUV, only include the FUV, or include both. {This test model used $f_{\textrm{PAH,d}}=1$ and grain growth}. The results of this exploration are given in Figure \ref{fig:EUVvsFUV}. Inclusion of the EUV has a negligible impact on the mass loss rate, at least until the disc is at $<0.03\,$pc from a $\theta^1$ Ori C type O star. When the FUV is included too, the EUV just sets the location of an ionisation front downstream in the flow. When only the EUV is included, there is a hot ($10^4\,$K), fast ($\sim10$\,km\,s$^{-1}$), but much more rarefied wind, meaning that the mass loss rate is lower. Ultimately for determining the mass loss rate and keeping calculations simple for users of the grid we opt not to include EUV radiation in \textsc{fried} at this stage.

{\cite{1998ApJ...499..758J} analytically studied EUV and FUV driven flows from discs. They found that the EUV dominates setting the mass loss very close to a $\theta^1$C type star. Moving away from the O star they infer that there will be an FUV dominated set of distances, but that then at larger distances still, the EUV will dominate again. The argument that the EUV dominates again at larger distances stems from their requirement that the FUV driven flows are super-critical (the outer disc is warmed above the escape velocity), which needs 1000\,K gas, or $>10^{4}$\,G$_0$. This means that discs could still be embedded within the H\,\textsc{ii} region beyond that threshold and the EUV radiation heats close to the disc surface. However, \cite{2004ApJ...611..360A} demonstrated that sub-critical winds can be launched down to weaker FUV fields (down to $\sim10^2$G$_0$), which is supported by other models \citep{2018MNRAS.481..452H} as well as observations in regions where the EUV is low such as \cite{2023arXiv230405777V}. It is unclear whether it is possible to have an FUV field $<10^2$\,G$_0$ within the photoionised gas of an H\,\textsc{ii} region but we suggest that generally if the EUV plays a dominant role it will only be in close proximity to the UV sources rather than at larger distances.  This is to be explored further in future work.}

{\cite{1998ApJ...499..758J} do provide an analytic expression for the EUV driven mass loss rate in section 2.3. If one wanted to consider the possible role of EUV driven mass loss the maximum of the \textsc{fried} and EUV driven mass loss rates could therefore be used, which is the approach used by \cite{2019MNRAS.490.5678C}. }

\subsection{Other improvements to the new grid}

\subsubsection{Capturing lower mass loss rates}
A disc subject to external photoevaporation is rapidly truncated until the mass loss rate drops to a value that matches the rate of viscous spreading \citep{2007MNRAS.376.1350C}. Accurate low mass loss rates are still therefore important for determining the disc radius once truncation has taken place and the external mass loss is less significant. An issue with the original \textsc{fried} grid is that it has a floor on the mass loss rate of $10^{-10}$\,M$_\odot$\,yr$^{-1}$. This was chosen because for weak winds the flow is sometimes pseudo-steady, making it difficult to determine the mass loss rate. The balance of viscous spreading and external photoevaporation leads to mass loss rates of the order of the accretion rate \citep{2020MNRAS.497L..40W}. Since accretion rates are observed down to $10^{-12}-10^{-10}$\,M$_\odot$\,yr$^{-1}$ \citep{2023ASPC..534..539M}, implementations of the existing grid have either had to terminate the model while it would still have observable accretion \citep{2020MNRAS.492.1279S}, or extrapolate below the floor \citep{2023arXiv230104656E, 2022A&A...666A..73B}. In the new version of the grid we {track the median mass loss rate in the flow over time} and impose no floor value to capture lower mass loss rates.

\begin{table*}
    \centering
    \begin{tabular}{l|c c l c}
         \hline
         \hline
    Parameter  & Symbol & Range  &  Notes  \\
    & & of values & \\
    \hline
    \hline
     Surface density at 1au &  $\Sigma_{1\textrm{au}}$   & $\left\{1, 10, 10^2, 10^3, 10^4, 10^5\right\}$\,g\,cm$^{-2}$ & \\
     Radius of disc boundary condition & $R_d$ & $\left\{5, 10, 20, 40, 60, 80, 100, 150, 250, 500\right\}$\,au & $\Sigma(R_d) = \Sigma_{1\textrm{au}}\left(\frac{R_d}{\textrm{au}}\right)^{-1}$ \\
     Stellar Mass & $M_*$ & $\left\{0.1, 0.3, 0.6, 1.0, 1.5, 3.0\right\}$\,M$_\odot$  \\
     PAH-to-dust ratio relative to ISM & $f_{\textrm{PAH,d}}$ & $\left\{0.1, 0.5, 1\right\}$ & $f_{\textrm{PAH,d}}=1$ gives ISM-like PAHs for ISM-like dust \\      
     UV field strength & $F_{\textrm{FUV}}$ & $\left\{1, 10, 10^2, 10^3, 10^4, 10^5\right\}$\,G$_0$  &  \\ 
     \hline
     \multicolumn{4}{c}{{Parameters associated with whether or not grain growth has occurred out to the disc outer edge} } \\
    Grain growth to disc outer edge?      & -- & Y/N  & Referred to as ``grain growth'' and ``ISM-like'' respectively\\
    UV cross section in the wind     & $\sigma_{\textrm{FUV}}$ & $4.5\times10^{-23}/8\times10^{-22}$cm$^2$ & Values from \cite{2016MNRAS.457.3593F} Figure 13 \\
    Dust-to-gas mass ratio in wind      & $\delta_{w}$ & $10^{-4}/10^{-2}$ &  \\
    Typical grain radius in wind      & $a_{w}$ &  $1$\,$\mu$m & Grain radius for grain surface chemistry  \\    
     Effective PAH-to-gas ratio relative to ISM & $f_{\textrm{PAH,g}}$ & $\left\{10^{-3}, 5\times10^{-3}, 10^{-2}\right\} / \left\{0.1, 0.5, 1\right\}$ &  \\     
    \hline
    \end{tabular}
    \caption{The parameters spanned by the \textsc{fried} v2 grid. It totals 12960 models comprised of 6 sub-grids with different microphysics. We consider the $f_{\textrm{PAH,d}}=1$ with grain growth subset to be the fiducial one (see section \ref{sec:fiducial} for discussion).  }
    \label{tab:FRIEDv2Parameters}
\end{table*}

\subsubsection{An update to the dust temperature calculation}
\label{sec:tdust}
The dust temperature in the UCL-PDR/3D-PDR code \citep{2005MNRAS.357..961B, 2006MNRAS.371.1865B, 2012MNRAS.427.2100B} which constitutes the underlying microphysics of these models is calculated following \cite{1991ApJ...377..192H}. This assumes that infrared emission re-radiated from the top of the PDR penetrates deeper into the PDR to set the dust temperature, but includes no attenuation of that infrared radiation. However, in dense PDRs like external photoevaporative winds this is not accurate. Attenuation of the infrared radiation has therefore been introduced by  using an approximation for the infrared-only dust temperature from \cite{1980ApJS...44..403R}              
\begin{equation}
    T_{\textrm{dust}} = T_0\left(N/N_0\right)^{-0.4}
\end{equation}                                                                 where $N_0$ is the column corresponding to $A_V=1$. This can lead to a cooler inner part of the flow in some cases, which reduces the mass loss rate compared to the original \textsc{fried} grid.

\section{The \textsc{fried} v2 parameter space}

In section \ref{sec:microphysics} we have determined that varying the base PAH-to-dust mass ratio and having the option for grain growth to have occurred or not adds the most utility to the \textsc{fried} grid. Metallicity variation, coolant depletion and the EUV could be included in future higher order expansions of the grid.  Adding these new features still dramatically increases the overall number of models required for the grid compared to \textsc{fried} v1. We therefore reduced the sampling in disc radius at large disc radii compared to the original \textsc{fried} grid. To validate this sparser sampling of macroscopic parameters we compared disc viscous evolutionary calculations with the original \textsc{fried} and a version of the original grid with sparser radial sampling beyond 100\,au, finding very close agreement. 

The parameters of the new grid are shown in Table \ref{tab:FRIEDv2Parameters}. The new grid totals 12960 models, more than triple the size of the original grid. In addition to the improvements discussed in section \ref{sec:microphysics}, we also extend the range of the FUV field. The lower limit is down from 10\,G$_0$ to 1\,G$_0$, a typical value in well-studied nearby regions like Lupus \citep{2016ApJ...832..110C}, and the upper limit up from $10^4$ to $10^5$\,G$_0$, which is at the level some proplyds are exposed to. This improves the situation where previously  users were having to extrapolate or use the mass loss rates at the boundaries of the grid. The surface density normalisation in the new grid spans 6 orders of magnitude, but the lowest surface density is actually slightly higher than in the previous grid. We checked this negligibly affects disc evolutionary models. The maximum disc radius is extended from 400\,au to 500\,au, whereas the minimum radius is increased from 1\,au to 5\,au. The increase in minimum radius is because the 1\,au values in the previous grid were basically all at the floor value, solutions there are very hard to obtain and external FUV driven mass loss from such a compact disc is not expected and internal winds will always dominate at those radii. 

Note that the only model that has an ISM-like PAH-to-gas ratio is $f_{\textrm{PAH,d}}=1$ with ISM-like dust (i.e. the case without dust growth in the disc). The other PAH-to-gas ratios are shown in the final row of Table \ref{tab:FRIEDv2Parameters} and are all lower, either due to a low dust-to-gas ratio or lower PAH-to-dust ratio. At this stage we consider the fiducial subset of the grid to be that with $f_{\textrm{PAH,d}}=1$ and grain growth, based on limited observational constraints on the PAH-to-gas ratio in propylds \citep[][]{2013ApJ...765L..38V} which we discuss further in section \ref{sec:fiducial}. 

For reference, the original \textsc{fried} grid had $f_{\textrm{PAH,d}}=0.1$, which was chosen to be conservative. The original grid assumed grain growth had happened in the disc, with a UV cross section in the wind of $\sigma_{\textrm{FUV}}=2.7\times10^{-23}$\,cm$^2$, whereas the models here with grain growth use $\sigma_{\textrm{FUV}}=4.5\times10^{-23}$\,cm$^2$ (i.e. there was less extinction in the wind for the original grid). The original \textsc{fried} grid assumed a dust-to-gas mass ratio in the wind of $\delta=3\times10^{-4}$, whereas in our dust depleted models we assume $\delta=10^{-4}$, which means that the PAH abundance in the original grid was a factor few higher than in our $f_{\rm PAH,d}=0.1$ models. These differences combine to mean that there is no perfect analogue to the original grid in the new grid.

We note that if the community has applications that \textsc{fried} v2 does not accommodate, smaller scale bespoke grids (e.g. for a single stellar mass regime and/or smaller range of other parameters) could be developed in collaboration by contacting the authors.

\section{Results: The \textsc{fried} v2 grid}

\subsection{The data release}
{Both the new and original versions of the \textsc{fried} grid are available as supplementary data to this paper, as well as on GitHub\footnote{https://github.com/thaworth-qmul/FRIEDgrid}}

The original \textsc{fried} grid was released as a single file with columns of: host star mass (M$_\odot$), FUV field strength (log$_{10}$\,G$_0$ ), surface density at the disc outer edge (log$_{10}$\,g\,cm$^{-2}$), disc (i.e. boundary condition) radius (au) and  mass loss rate (log$_{10}$\,M$_\odot$\,yr$^{-1}$). Rather than distribute the new grid as a single file, we release it as 36 files for different stellar masses, $f_{\textrm{PAH,d}}$ values and dust parameters. They are named as, for example 
\begin{verbatim}
    FRIEDv2_0p1Msol_fPAH0p1_growth.dat
\end{verbatim}
where "0p1Msol" refers to the host star mass being 0.1\,M$_\odot$, "fPAH0p1" refers to $f_{\textrm{PAH,d}}=0.1$ and "growth" referring to this being for the case of grain growth in the disc (the alternative being "{ISM}"). The columns in the new files are then: host star mass (M$_\odot$), disc (i.e. boundary condition) radius (au), the assumed surface density normalization $\Sigma_{1\textrm{au}}$ (g\,cm$^{-2}$), the surface density at the disc outer edge $\Sigma(R_d)$ (g\,cm$^{-2}$), the incident FUV radiation field strength (G$_0$) and the mass loss rate (log$_{10}$\,M$_\odot$\,yr$^{-1}$). We  provide advice on incorporating these mass loss rates into disc viscous evolutionary calculations in section \ref{sec:lessons}. 

\begin{figure*}
    \centering
    \includegraphics[width=1.0\columnwidth]{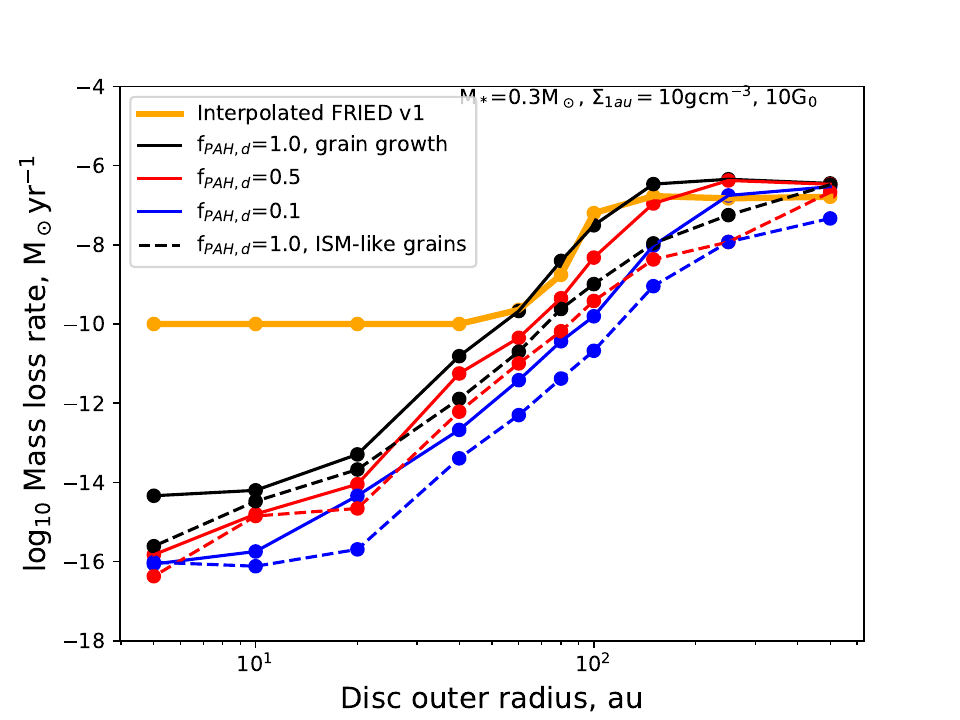}
    \includegraphics[width=1.0\columnwidth]{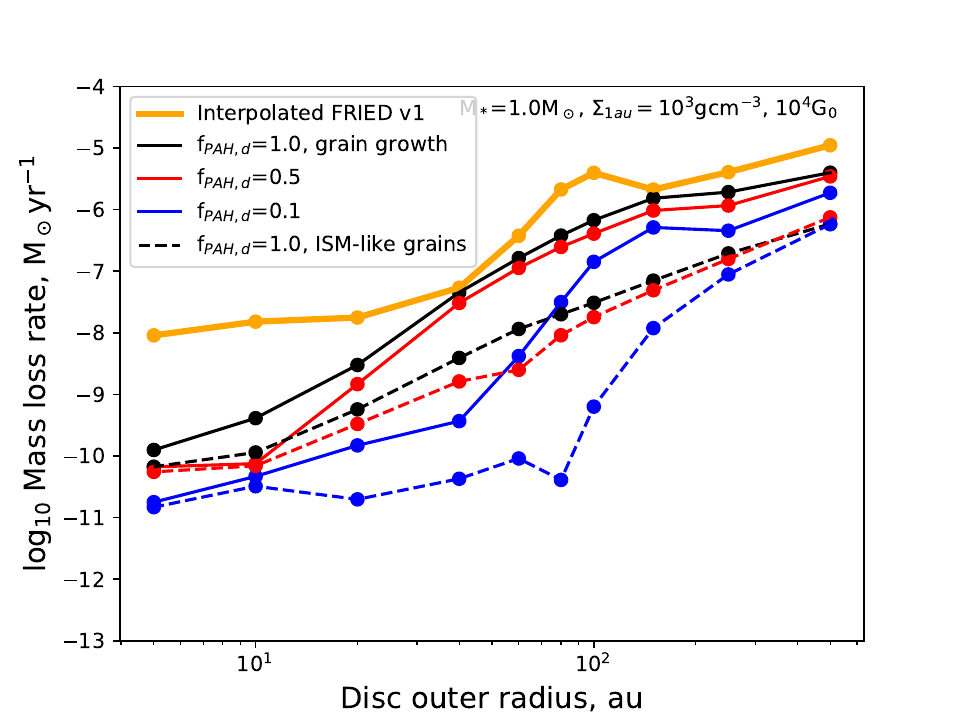}

    \includegraphics[width=1.0\columnwidth]{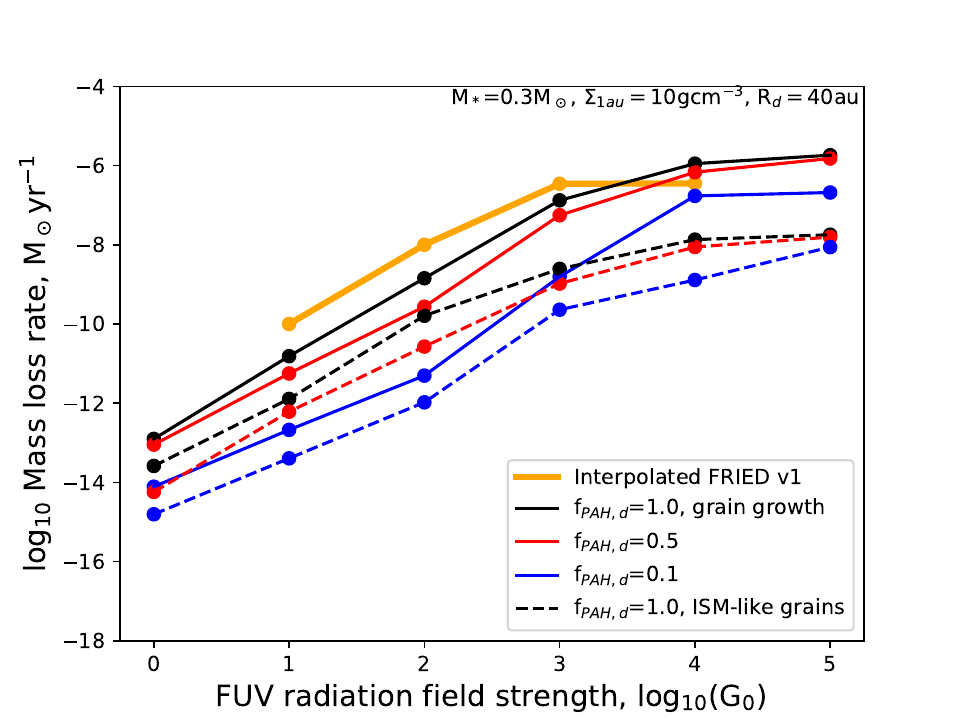}
    \includegraphics[width=1.0\columnwidth]{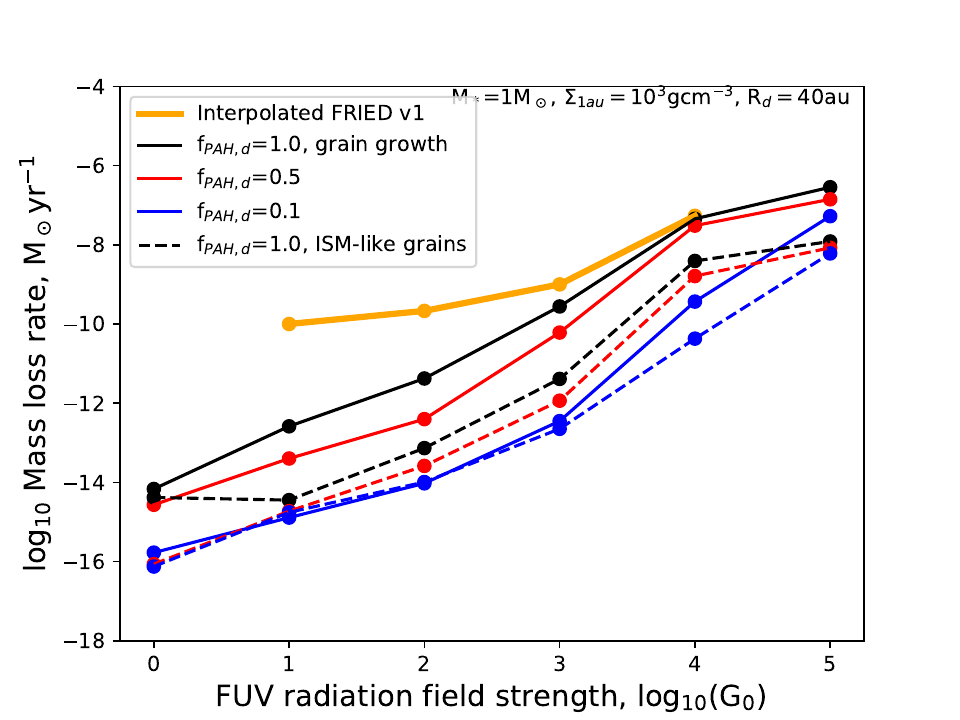}
    \caption{Example \textsc{fried} v2 mass loss rates as a function of disc outer radius (upper panels) and as a function of the external UV radiation field (lower panels). The properties of each model are defined in the upper right of each panel. The solid lines are models where grain growth has taken place in the disc outer regions, whereas the dashed lines still have small dust in the outer disc. The black, red and blue lines are for $f_{\textrm{PAH,d}}=1, 0.5$ and 0.1 respectively. The orange lines are interpolated from the original \textsc{fried} grid. } 
    \label{fig:V1_comparison}
\end{figure*}

\subsection{A look at a subset of the grid}
The \textsc{fried} v2 grid is sufficiently large that it is difficult to discuss in its entirety in detail. However we explore and discuss a small subset to illustrate some key behaviours here and in Figures \ref{fig:PAH1p0grow}--\ref{fig:PAH0p1small} in the appendix include a larger set of plots for reference.

Figure \ref{fig:V1_comparison} shows the mass loss rate as a function of radius in the upper panels and as a function of the external UV radiation field in the lower panels. In each panel the black, red and blue lines correspond to $f_{\textrm{PAH,d}}=1, 0.5$ and $0.1$ respectively. Solid lines assume grain growth, whereas dashed {is ISM-like dust} in the outer disc. The orange lines are interpolated from the original \textsc{fried} grid for comparison. The left hand panels are a 0.3\,M$_\odot$ star with $\Sigma_{1au}=10$\,g\,cm$^{-3}$. The upper left has an external FUV radiation field of 10\,G$_0$, the lower left has a disc outer radius of 40\,au. The right hand panels are a 1\,M$_\odot$ star with $\Sigma_{1au}=10^{3}$\,g\,cm$^{-3}$. The upper right has an external FUV radiation field of 10$^4$\,G$_0$ and the lower right has a disc outer radius of 40\,au.

There is no perfect analogue here to the original \textsc{fried} models, though those are closest to the models assuming grain growth (solid lines), which also assumed in \textsc{fried} v1. Generally though, if anything the old grid has higher mass loss rates. The general behaviour is that as the PAH abundance is reduced the mass loss rate decreases. There is also a significant drop in mass loss rate if the outer disc still has small dust (dashed lines). We will briefly explore the importance of this in section \ref{sec:discussion}, but the lower mass loss rates early on could be important for protecting the planet forming reservoir. 

{All panels in Figure \ref{fig:V1_comparison} show the importance of overcoming the old floor value of $10^{-10}$\,M$_\odot$\,yr$^{-1}$, and that the new grid accurately captures mass loss rates down to at least 6 orders of magnitude lower than the old floor value.  }

The general and well known behaviour of the mass loss rate with disc radius (smaller discs are more bound and so have lower mass loss rates) and external UV field strength (higher UV fields lead to higher mass loss rates) is maintained here. 

These examples also show that the new grid, which extracts mass loss rates from the models entirely automatically, is not perfectly ``noise free'' (nor was the original grid). This noise is at a sufficently low level that the different microphysical regimes are distinct. For example for a given $f_{\textrm{PAH,d}}$ whether grain growth has occurred or not regularly changes the mass loss rate by an order of magnitude. {We note that the feature at 100\,au in the upper right hand panel is real (as in it is not noise) and also appears in some other models. It is a result of the mass loss rate being sensitive to both the surface density at the disc outer edge (which is a function of disc radius) and the disc outer radius.}

\begin{figure*}
    \centering
    \includegraphics[width=\columnwidth]{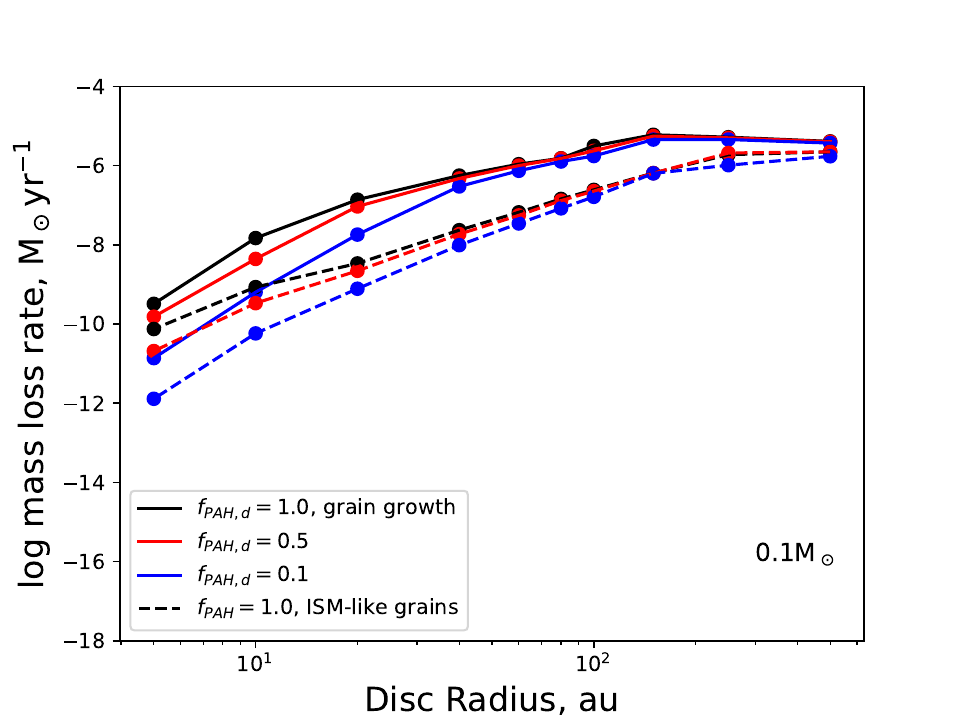}
    \includegraphics[width=\columnwidth]{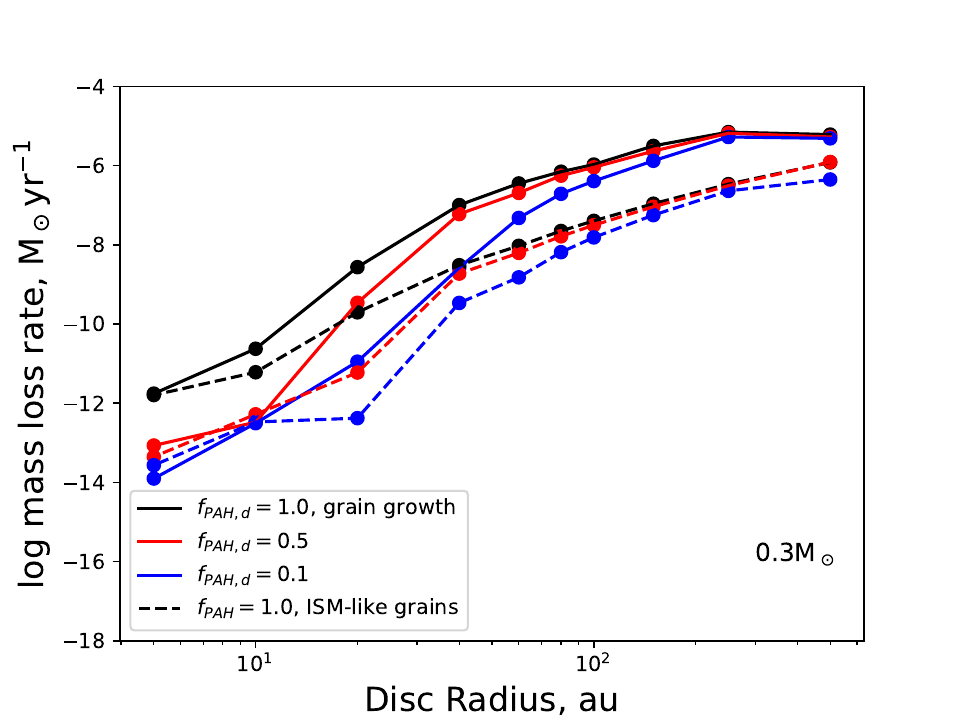}

    \includegraphics[width=\columnwidth]{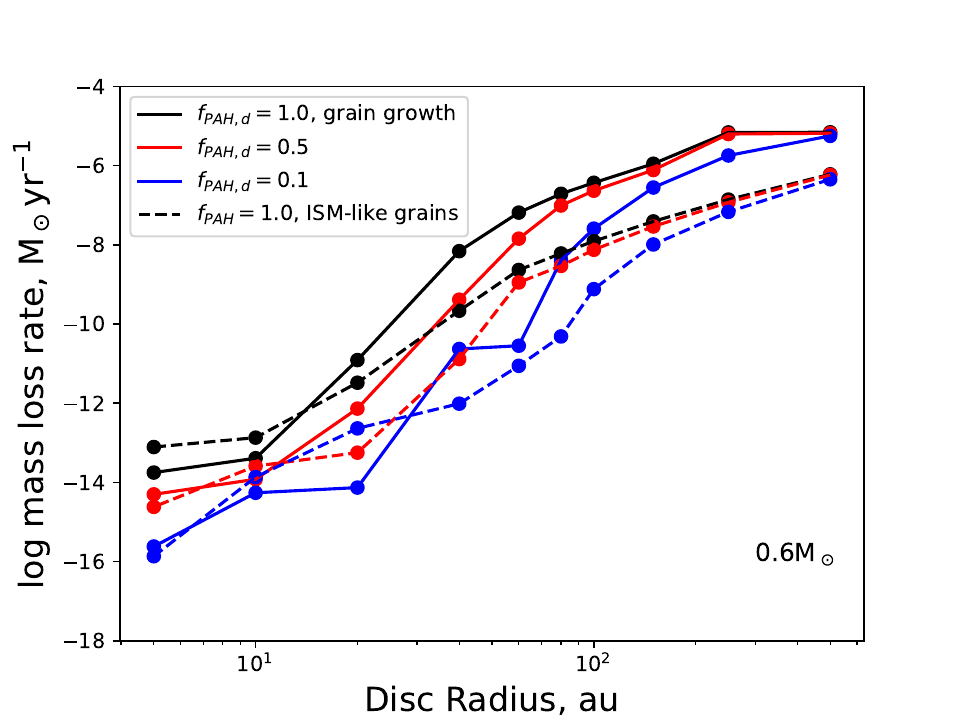}
    \includegraphics[width=\columnwidth]{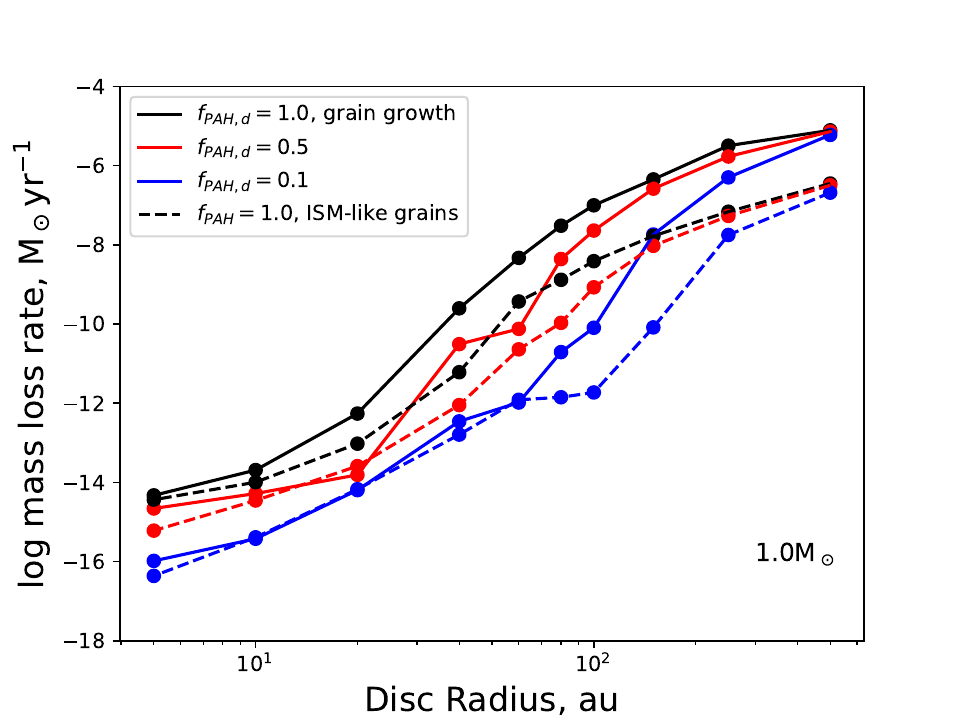}

    \includegraphics[width=\columnwidth]{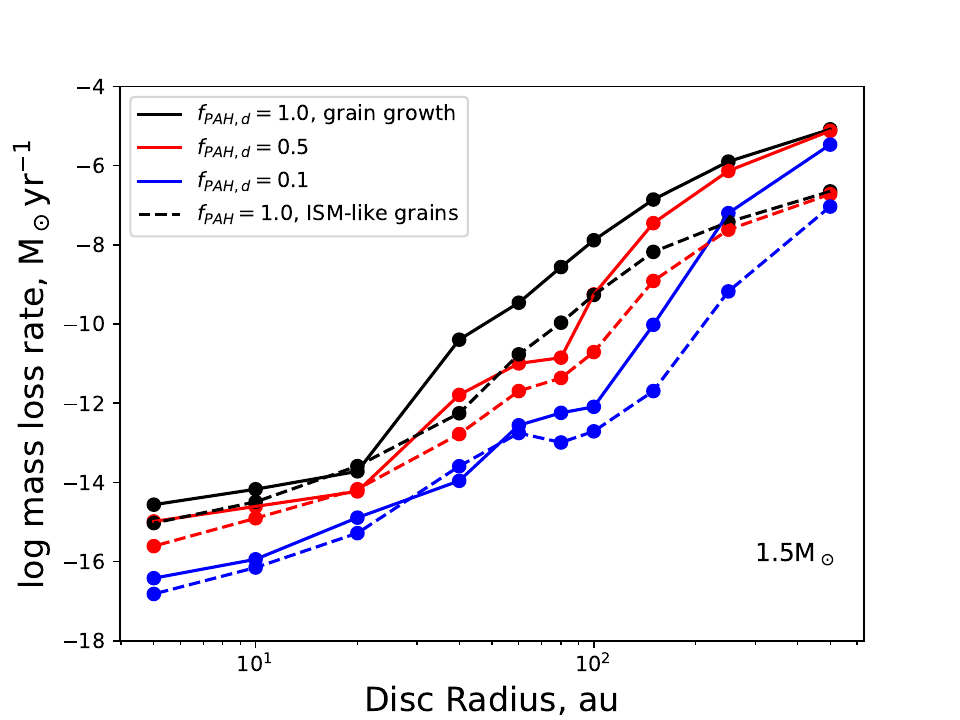}
    \includegraphics[width=\columnwidth]{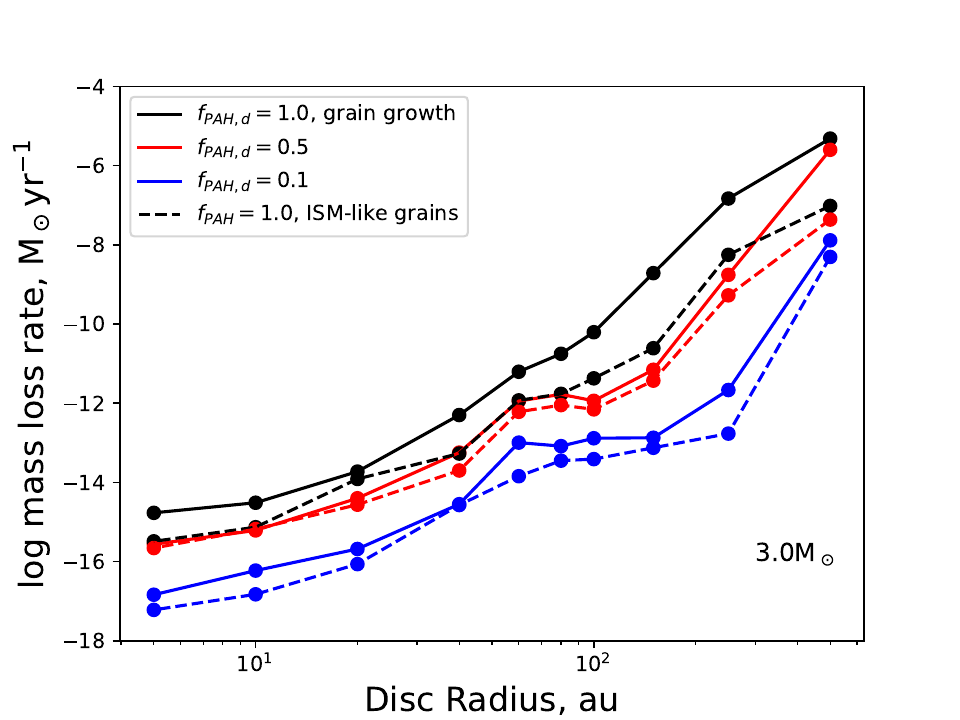}
    \caption{Mass loss rate as a function of disc outer radius for an external UV field of $10^3$\,G$_0$, $\Sigma_{1au}=10^2$\,g\,cm$^{-2}$ and host star mass of 0.1, 0.3, 0.6, 1.0, 1.5 and 3.0M$_\odot$ from left to right, top to bottom. }
    \label{fig:varyMass}
\end{figure*}

Figure \ref{fig:varyMass} shows some additional examples for host star masses of 0.1, 0.3, 0.6, 1.0, 1.5 and 3\,M$_\odot$ from left to right, top to bottom. There is the same general sensitivity to the microphysics and disc radius parameters as discussed for Figure \ref{fig:V1_comparison}. As expected from previous work, as the host star mass increases the boundedness of the outer disc increases and the mass loss rates decreases.

\begin{figure}
    \centering
    \includegraphics[width=\columnwidth]{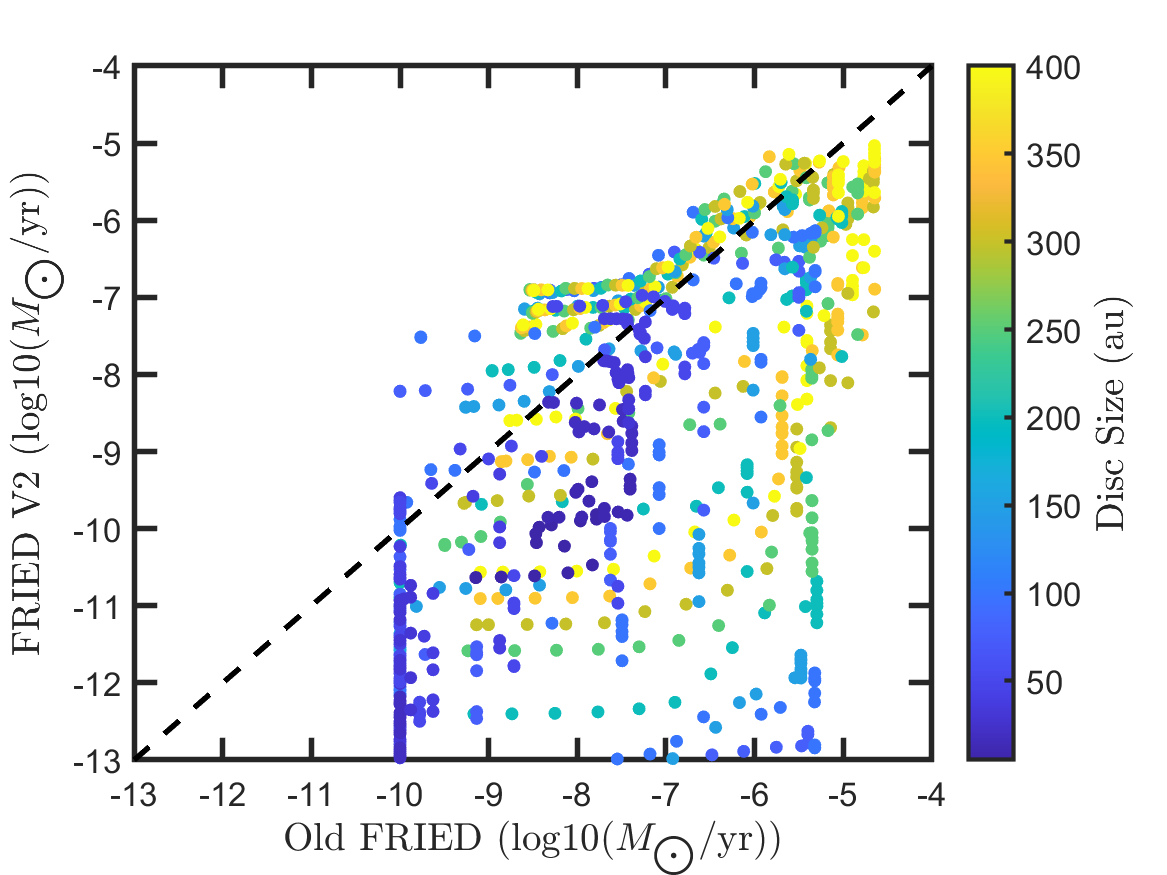}
    \caption{A comparison of the original \textsc{fried} grid mass loss rates with those from the $f_{PAH}=1$, grain growth models of the new \textsc{fried} grid presented here. The colour scale denotes the disc radii and the host star mass in all cases is 1\,M$_\odot$. This plot illustrates that typically the new grid has lower mass loss rates. The feature at $\textrm{log}_{10}{\dot{M}}=-10$ is due to the floor value in the v1 grid. }
    \label{fig:v1v2comparison}
\end{figure}

Figure \ref{fig:v1v2comparison} compares the mass loss rates in the new \textsc{fried} grid with $f_{\textrm{PAH,d}}=1$ and grain growth (which in section \ref{sec:fiducial} we suggest is the fiducial one) with interpolated values from the original grid in the 1\,M$_\odot$ host star case. The points are colour coded by disc radius. Figure \ref{fig:v1v2comparison} illustrates that there are regions of the parameter space that are consistent, but also components where the new grid gives higher, and others lower, mass loss rates. The abrupt feature at $\textrm{log}_{10}{\dot{M}}=-10$ is due to the floor value in the original grid. 

\begin{figure}
    \centering
    \includegraphics[width=\columnwidth]{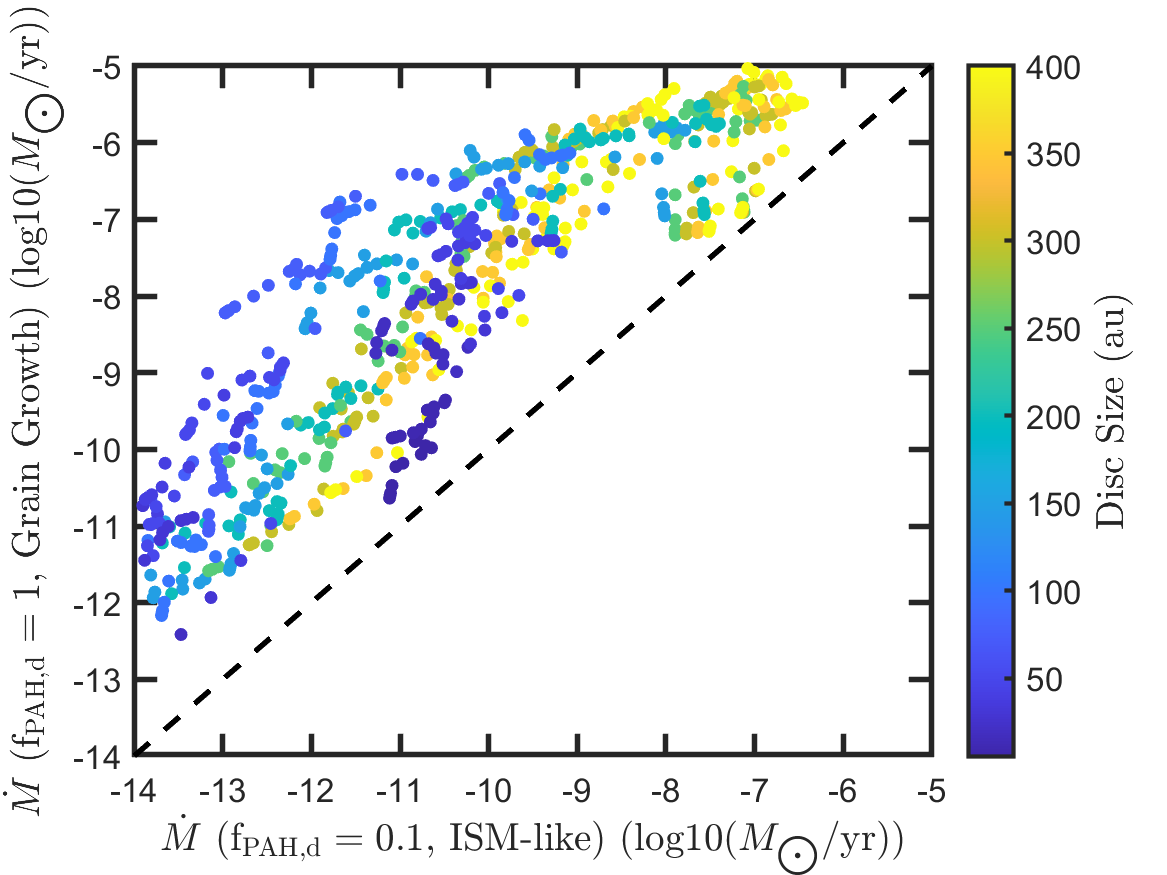}
    \caption{A comparison of the two most different subsets of the new \textsc{fried} grid. The vertical is mass loss rates in the case of grain growth and $f_{\textrm{PAH,d}}=1$ (typically giving the highest mass loss rates) and horizontal is the {ISM-like dust} case with $f_{\textrm{PAH,d}}=0.1$ (typically giving the lowest mass loss rates). }
    \label{fig:Extremes}
\end{figure}

Finally Figure \ref{fig:Extremes} compares the mass loss rates in the new grid in the case with grain growth and $f_{\textrm{PAH,d}}=1$, which gives the highest mass loss rates, and the case with {ISM-like dust} and $f_{\textrm{PAH,d}}=0.1$, which gives the lowest mass loss rates. This highlights that the choice of sub-grid can affect the mass loss rates by orders of magnitude, which underscores the importance of empirically determining the PAH abundance in discs (particularly in externally photoevaporating discs), and the  possible role of small dust in reducing the impact of external photoevaporation at early times before the dust grows to sizes that is not entrained in the wind. 

\section{Lessons learned: some hints regarding implementing \textsc{fried} in disc evolutionary models}
\label{sec:lessons}
Many different groups have utilised the original \textsc{fried} grid in  disc evolutionary calculations and a lot has been learned about the challenges in the implementation, and means of overcoming them. Here we provide a brief overview of some of the known challenges and ways of resolving them. 

In 1D  external photoevaporation models like \textsc{fried}, external photoevaporation extracts mass from the outermost part of the disc only. In 2D (or 3D) discs do not exclusively lose mass from the disc outer edge, but the mass loss is genuinely from the outer 20\,per cent of the disc or so \citep{2019MNRAS.485.3895H}. Either way, the conditions in the outer disc are important.

\subsection{Where is the disc outer edge?}
\label{sec:rd}
The main note of caution that has arisen in implementing \textsc{fried} is that to obtain a mass loss rate a disc outer radius and surface density is required. Discs typically have an exponential fall off in surface density in the outer disc, and the mass loss rate is sensitive to both radius and surface density. So there is an issue in that an incremental change in the chosen ``disc radius'' in this exponential tail can lead to a large change in surface density and hence mass loss rate. So how do we define the outer radius for a continuous disc in a manner that is robust? {In reality the wind will be driven close to the $\tau=1$ surface, however if the wind profile is not being solved for (as is the case in these viscous evolutionary models) that cannot be trivially calculated.  }

Resolving this potential issue has been done in a number of ways, all of which have provided qualitatively similar results. For example \cite{2019MNRAS.490.5678C} define a floor surface density of $10^{-12}$\,g\,cm$^{-2}$ and consider the disc outer edge to be the cell adjacent to where that floor value is reached {(though note that that will not reliably identify the $\tau=1$ surface)}. Another example is that \cite{2020MNRAS.492.1279S} identify the point at which the \textsc{fried} grid mass loss rates have optically thick solutions. They do so by calculating the \textsc{fried} mass loss rate at all radii in the disc, defining the disc outer radius (and hence relevant mass loss rate) as the location of maximum mass loss rate. {When the wind becomes optically thin, the velocity is independent of the density normalisation and so the mass-loss rate becomes linear in the density \citep{2016MNRAS.457.3593F}. Thus a strongly declining mass-loss rate with radius is a hallmark that the wind would be optically thin if launched there. The maximum therefore gives an approximation to the $\tau=1$ surface where these optical depth effects dominate over the effects of the gravitational potential.} They then compute a weighted mass loss rate from that point of maximum mass loss rate outward through the disc \citep[see][for full details]{2020MNRAS.492.1279S}. We suggest new users study the implementation in these papers. { A further benefit of this approach is that since it relies on the intrinsic properties of an optically thin flow, it should consistently approximate the $\tau=1$ surface regardless of the microphysics.}

The points above relate to calculating the mass loss rate. We also caution briefly that the disc outer edge in \textsc{fried} does not necessarily correspond to an observed gas disc radius (e.g. a CO radius). In particular the CO dissociation front can often be in the wind \citep{2018MNRAS.481..452H}. A good example of why this is IM Lup. That disc has CO emission out to about 1000\,au, but exhibits a break in the surface density profile at $\sim400$\,au \citep{2009A&A...501..269P}. \cite{2017MNRAS.468L.108H} found that a CO extent of around 1000\,au could be reproduced in an external photoevaporative wind model where the "disc outer radius" in the language of a \textsc{fried}-type calculation was 400-450\,au. IM Lup is an extreme example in a low UV environment. At higher UV field strengths the CO dissociation front does move closer to the ``disc outer edge''.

\subsection{What should be considered the fiducial \textsc{fried} v2 subset?}
\label{sec:fiducial}
\textsc{fried} has been widely used, and here we introduce additional dimensionality that users may not be familiar with. In particular, some subsets of the new grid have substantially lower mass loss rates. There is therefore a danger that this exploration of parameters ends up being misinterpreted. We hence discuss here which subsets we consider to be fiducial, and which are included as more of an exploratory tool given the lack of observational constraints on the microphysics parameters. 

The new parameters are whether or not grain growth has occurred, and the PAH-to-dust ratio. Grain growth to sizes that are not entrained in the wind proceeds reasonably quickly in the disc, from the inside-out. It is for that reason that the original grid considered a dust-depleted wind. However for a 100\,au disc it could take $\sim$1\,Myr to move to the {``grain growth''} state. The lower mass loss rates in that period predicted by \textsc{fried} v2 could help with solids retention for planet formation (we leave it to the community to explore this). 

There are very limited constraints on the PAH abundance in external photoevaporative winds, which is why we include a substantial range here. As facilities like JWST provide further constraints, use of the grid can adapt to the findings of those studies. {So far there is just one estimate of the PAH-to-gas ratio $f_{\textrm{PAH,g}}$, which is for the proplyd HST 10.  For that proplyd \cite{2013ApJ...765L..38V} inferred a PAH-to-gas value that is 1/50th that of the surrounding Orion Bar and 1/90th that in NGC 7023 \citep{2012PNAS..109..401B}. Overall, while we have some idea that the PAH-to-gas abundance is probably slightly depleted, we don't have a good idea of a fiducial value, nor to what extent this is due to PAH depletion or dust depletion. }

The PAH-to-gas abundance in the 6 subsets of the new \textsc{fried} grid are
\begin{itemize}
    \item $f_{\textrm{PAH,d}}=1$, {ISM-like dust}: PAH-to-gas is $f_{\textrm{PAH,g}}=1 $
    \item $f_{\textrm{PAH,d}}=0.5$, {ISM-like dust}: PAH-to-gas is $f_{\textrm{PAH,g}}=1 /2$
    \item $f_{\textrm{PAH,d}}=0.1$, {ISM-like dust}: PAH-to-gas is $f_{\textrm{PAH,g}}=1 /10$
    \item $f_{\textrm{PAH,d}}=1$, grain growth: PAH-to-gas is $f_{\textrm{PAH,g}}=1 /100 $
    \item $f_{\textrm{PAH,d}}=0.5$, grain growth: PAH-to-gas is $f_{\textrm{PAH,g}}=1 /200$ 
    \item $f_{\textrm{PAH,d}}=0.1$, grain growth: PAH-to-gas is $f_{\textrm{PAH,g}}=1 /1000$     
\end{itemize}
With \cite{2013ApJ...765L..38V} finding PAH depletion at the level of 1/50 -- 1/90, \textit{At this stage we therefore consider the fiducial subset of the grid to be that with $f_{\textrm{PAH,d}}=1$} and grain growth, which has a PAH-to-gas ratio of $f_{\textrm{PAH,g}}=1/100$. This would also imply that the {ISM-like dust} version with $f_{\textrm{PAH,d}}=1$ would be fiducial until the pebble production front reaches the disc outer edge (how one transitions from the {ISM-like dust} to grain growth subsets {of the grid} could conceivably be done in various ways and is left to the community to explore).  

So our advice is using $f_{\textrm{PAH,d}}=1$ as the fiducial value, however $f_{\textrm{PAH,d}}=0.5$ is certainly plausible and further observational constraints are required to pin down the microphysics of external photoevaporation. Users are free to explore the impact of different subsets on disc evolution and planet formation. That said, we wouldn't advise, for example, drawing strong conclusions from models exclusively using the $f_{\textrm{PAH,d}}=0.1$ {ISM-like dust} case which has a corresponding PAH-to-gas ratio of $f_{\textrm{PAH,g}}=1/1000$ (orders of magnitude lower than our existing constraints). 

\subsection{Using the grid efficiently}
A more minor problem encountered in the use of the original \textsc{fried} is that it is computationally expensive to interpolate over a four-dimensional $\dot{M}(\Sigma_d, R_d, G_0, M_*)$ grid. However for many applications this has been circumvented by using a lower dimensionality subset of the grid. For example, removing the sensitivity to stellar mass for a disc evolving around a given star \citep{2022MNRAS.512.3788Q}, which is reasonable because the stellar mass does evolve slowly compared to the timescale for external photoevaporation \citep[e.g.][]{2016ApJ...823..102C, 2021MNRAS.508.3710R}. Others assume a constant external radiation field \citep[e.g.][]{2018MNRAS.478.2700W, 2020MNRAS.492.1279S}. 

These reductions in dimensionality are valid and can be continued with the new version of the grid. In particular it makes sense to use a subset of the grid (or a new interpolated grid) for a single stellar mass in any given calculation. The UV field may vary substantially in time \citep{2022MNRAS.512.3788Q, 2023MNRAS.520.5331W, 2023MNRAS.520.6159C} but fixing is still reasonable for some applications. For the microphysics, we would generally advise either using a single subset of the grid at a time rather than interpolating between them (e.g. a single $f_{\textrm{PAH,d}}$ value). If interested in following how the external photoevaporation changes as  dust evolves, we would advise using a version of the grid with {ISM-like dust} in the outer disc until the pebble production front  reaches the disc outer edge, then switching entirely to the subset that considers grains to have grown, and small dust to be depleted in the outer disc.

\section{Discussion}
\label{sec:discussion}

\subsection{An initial comparison of disc viscous evolution}
\label{sec:discevo}
We now make an initial exploration of how discs evolve in different components of the new \textsc{fried} grid compared to the old. There are myriad science applications with these kind of models, but our goal here is simply to illustrate how different the forms of the grid are. 

We use the disc evolutionary code of \citet{2021MNRAS.506.3596C} that has previously been used for studies of the evolutionary pathways of internally and externally photoevaporating discs \citep{2022MNRAS.514.2315C}. This code was adapted to include the new \textsc{fried} grid and interpolate within it to calculate mass loss rates following \citet{2020MNRAS.492.1279S,2022MNRAS.512.3788Q}. The reader should refer to these papers for full details of the methodology. However in short, these are viscous disc evolutionary calculations with accretion and a pseudo viscous $\alpha=10^{-3}$. They include no internal winds, but extract mass from the outer regions of the disc due to external photoevaporation. We consider solar mass stars with discs of initial mass of 0.1\,M$_{\odot}$ and a $R^{-1}$ surface density profile with an exponential cutoff beyond 50\,au. We only consider the gas evolution in this paper and leave consideration of the dust evolution to future work. We also reiterate that this is just one set of star-disc parameters and that the detailed evolution of discs is going to be sensitive to those. 

\begin{figure*}
    \centering
    \includegraphics[width=2\columnwidth]{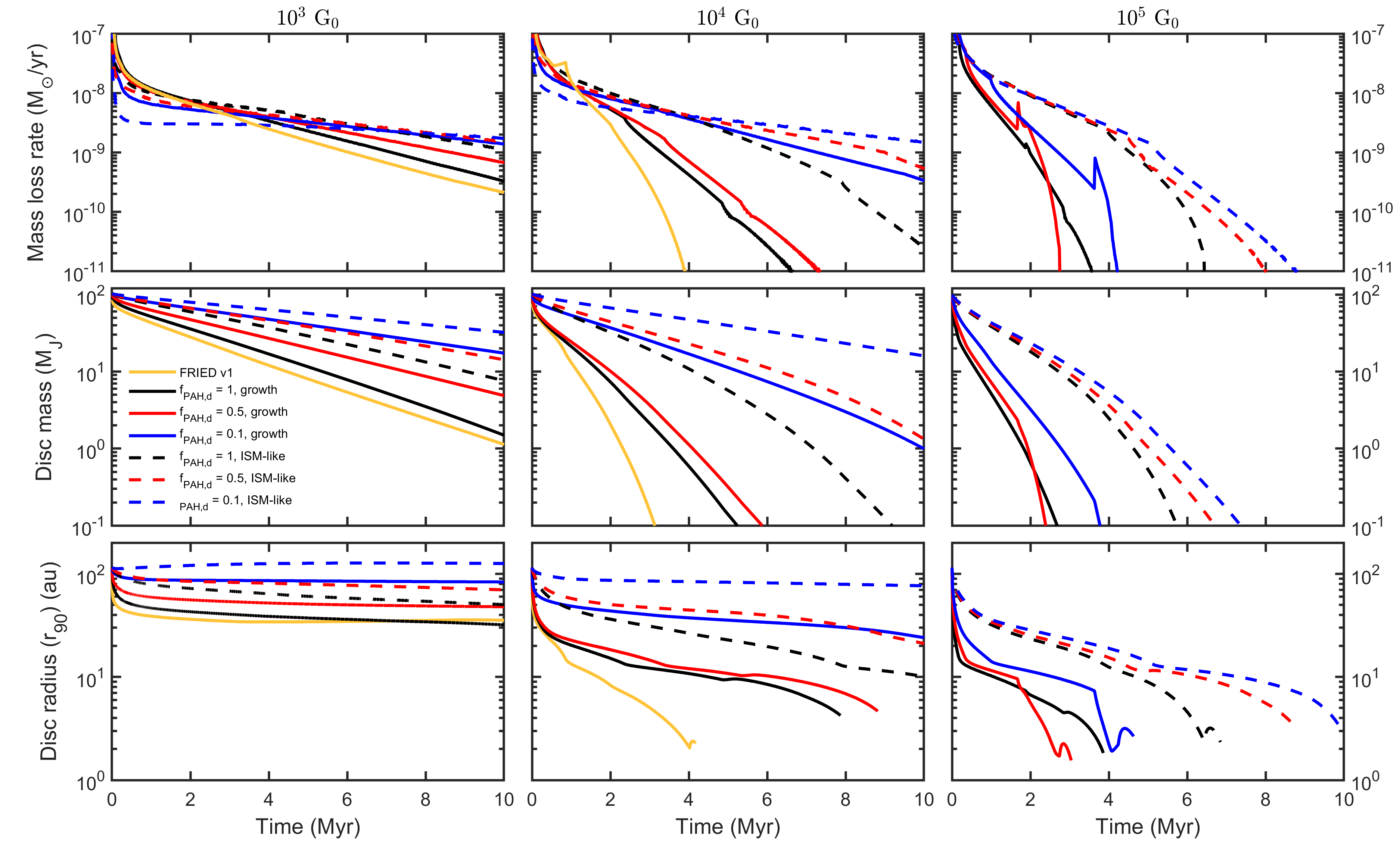}
    \caption{Illustrative disc evolutionary calculations using the new \textsc{fried} grid. The top panels show the temporal evolution of mass loss rates, the middle panels show the evolution of disc masses, whilst the bottom panels show the evolution of the radius in the disc containing 90\,per cent of the mass. These are all for the case of a solar mass star with a disc that is initially 0.1\,M$_\odot$ and  has an exponential cutoff in the disc surface density profile at 50\,au. }
    \label{fig:disc_evol}
\end{figure*}
 
In fig. \ref{fig:disc_evol} we show the temporal evolution of mass loss rates (upper panels), disc mass (middle panels) and disc radii (lower panels). The FUV radiation field is $10^3, 10^4$ and $10^5$\,G$_0$ from left to right. Note that here we define the disc radius as that containing 90\,per cent of the mass. In the left and central columns the orange line shows the evolution of a model utilising the original \textsc{fried} grid. The original grid did not go to $10^5$\,G$_0$, so there is no orange line in the right hand column, though we note that in previous applications such as \cite{2022MNRAS.512.3788Q} the $10^4$\,G$_0$ mass loss rates were used wherever the UV field strength exceeded that upper limit, so for a comparison against what has been done in the past, one could compare the behaviour of the orange lines in the central column with those in the right column. In each panel the other lines follow the same colour and linestyle used throughout the rest of the paper. Models with $f_{\rm PAH,d}=1, 0.5, 0.1$ are black, red and blue respectively. Solid lines are with grain growth (a dust depleted wind) and dashed lines are with {ISM-like dust} in the outer disc. 

There are many interesting features in this collection of disc evolutionary calculations. In the left hand column the orange and black lines (original grid and the new fiducial model with dust depletion)  are fairly consistent (note that these models do have different microphysics). In the central column both of these models are rapidly ($<1\,$Myr) truncated down to $\sim20$\,au, however there they diverge, with the model using the original grid being completely dispersed more quickly. So there is no qualitative change in the expectation of rapid truncation, but the longevity of inner discs (at least without internal winds) appears to be increased in the new grid. 

The different subsets of the new grid behave in the way one would expect based on the discussion of the mass loss rates above. Grain growth in the disc (i.e. dust depleted winds) lead to higher mass loss rates and hence more rapid truncation of the disc than those with {ISM-like dust}. Similarly reducing the PAH-to-dust ratio decreases the mass loss rate and leads to less rapid disc truncation.

The right hand column shows disc evolution in a UV field strength beyond the upper limit of the original \textsc{fried} grid. We note that in that right hand column all calculations with grain growth see the disc completely dispersed on timescales shorter than the original grid would have concluded (since it was capped at $10^4$\,G$_0$), irrespective of the PAH abundance. So while the new grid quite widely introduces lower mass loss rates, in high UV environments such as those typically encountered by proplyds the new grid results in higher mass loss rates.

\subsection{Comparison with proplyds}

Proplyds have mass loss rate estimates computed empirically based on the extent of the ionisation front and ionisation balance arguments \cite[see][]{1999AJ....118.2350H, 2021MNRAS.501.3502H, 2022EPJP..137.1132W}. This routinely yields mass loss rates $10^{-7}-10^{-6}$\,M$_\odot$\,yr$^{-1}$ (though note the proplyds in the inner ONC are exposed to radiation fields more like $10^6$\,G$_0$ and are in a regime where the EUV really does probably play a role). Those high mass loss rates compared to typical disc lifetimes leads to the idea of the proplyd lifetime problem, because discs should be dispersed so fast that we wouldn't expect to see them \citep{1999AJ....118.2350H}. A possible solution to this is ongoing star formation and stellar motions in the cluster \citep{2019MNRAS.490.5478W}. 

The evolutionary calculations in the right hand column of Figure \ref{fig:disc_evol} are interesting in comparison with proplyds. On the one hand the models with {ISM-like dust} entrained in the wind live substantially longer, so dust entrainment could help alleviate some of the proplyd lifetime problem. However it cannot be the entire solution since the high mass loss rates inferred are empirical, and the mass loss rates in the long-lived models with {ISM-like dust} are much lower. This underscores the importance of understanding and observationally constraining the nature of grain entrainment in external photoevaporative winds, and understanding how long the transition to being dust depleted takes. 

\section{Summary}
We introduce and explore a new \textsc{fried} grid of mass loss rates from protoplanetary discs due to external photoevaporation. External photoevaporative winds are driven from the outer regions of discs when nearby massive stars expose them to significant amounts of UV radiation. This new grid expands on the original version of \textsc{fried} \citep{2018MNRAS.481..452H} in terms of the breadth of parameter space in UV field, stellar mass, disc mass and disc radius. The new grid also provides the option to use different PAH-to-dust ratios, which is important because PAH's can provide the main heating mechanism in PDRs and their abundance is uncertain. Finally, the new grid provides the option to control whether or not grain growth has happened in the disc, which affects the opacity in the wind since only small grains are entrained. We find that these are the most important microphysics considerations, with the overall metallicity, inclusion of EUV and coolant depletion being less important. There are also additional improvements, including removing a floor value in the possible mass loss rates, more robust mass loss rate estimates and checks that the compact discs in weak UV environments (where numerically stable solutions are most difficult to compute) do not violate energy limited considerations. Overall the new grid contains 12960 models. We also undertake an initial exploration of the disc evolutionary calculations with different external photoevaporation microphysics.

The mass loss rates range from being similar to the original grid to being lower, depending on the microphysics. In particular, only small dust grains are entrained in external photoevaporative winds, so whether or not grain growth has taken place in the disc can dramatically influence the extinction in the wind and hence the UV incident on the disc, the mass loss rate and the resulting disc evolution. 

What we consider to be the fiducial subset of the grid is that with a PAH-to-dust ratio $f_{\textrm{PAH,d}}=1$ with grain growth in the disc. This subset has a PAH-to-gas ratio depleted by a factor 100, and the proplyd HST 10 is observed to have a PAH-to-gas ratio depleted by a factor 50 -- 90 \citep[][]{2013ApJ...765L..38V}. In this framework, the reduced PAH-to-gas abundance is due to a reduction in the dust-to-gas ratio in the wind. Despite the aforementioned being the fiducial, we expect that at early times the model without grain growth in the disc will be more valid (the exact timescale is dependent on disc radius, but typically $<1$\,Myr), after which the model with grain growth is most applicable. Despite our arguments above, the PAH abundance in the wind is not generally well understood so other subsets of the grid with lower PAH abundance can be used to explore the possible implications for disc evolution and planet formation, and may transition to being more fiducial as further constraints on the PAH abundance are obtained.


\section*{Data availability}
The \textsc{friedv2} and original \textsc{fried} grids of mass loss rates are made publicly available as online supplementary data associated with this paper.

\section*{Acknowledgements}
We thank the community for their use of \textsc{fried}, and for their support and feedback. We encourage further engagement in the future. 
We thank Shmuel Bialy for sharing the results that we used for benchmarking the metallicity variation. 
TJH acknowledges funding from a Royal Society Dorothy Hodgkin Fellowship and UKRI guaranteed funding for a Horizon Europe ERC consolidator grant (EP/Y024710/1). 
This work was performed using the DiRAC Data Intensive service at Leicester, operated by the University of Leicester IT Services, which forms part of the STFC DiRAC HPC Facility (www.dirac.ac.uk). The equipment was funded by BEIS capital funding via STFC capital grants ST/K000373/1 and ST/R002363/1 and STFC DiRAC Operations grant ST/R001014/1. DiRAC is part of the National e-Infrastructure.
This research utilised Queen Mary's Apocrita HPC facility, supported by QMUL Research-IT (http://doi.org/10.5281/zenodo.438045).
 
\bibliographystyle{mnras}
\bibliography{molecular}

\appendix

\section{The new \textsc{fried} grid. }
\begin{figure*}
    \includegraphics[width=2.0\columnwidth]{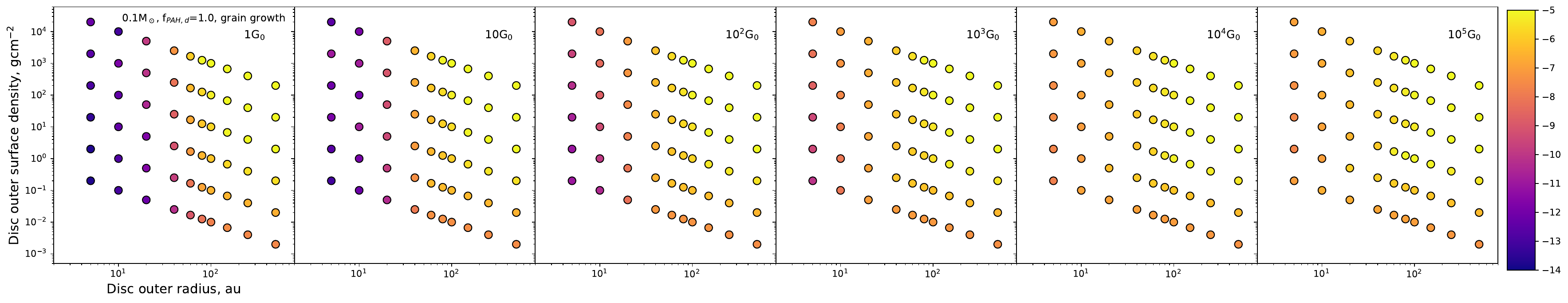}
    \includegraphics[width=2.0\columnwidth]{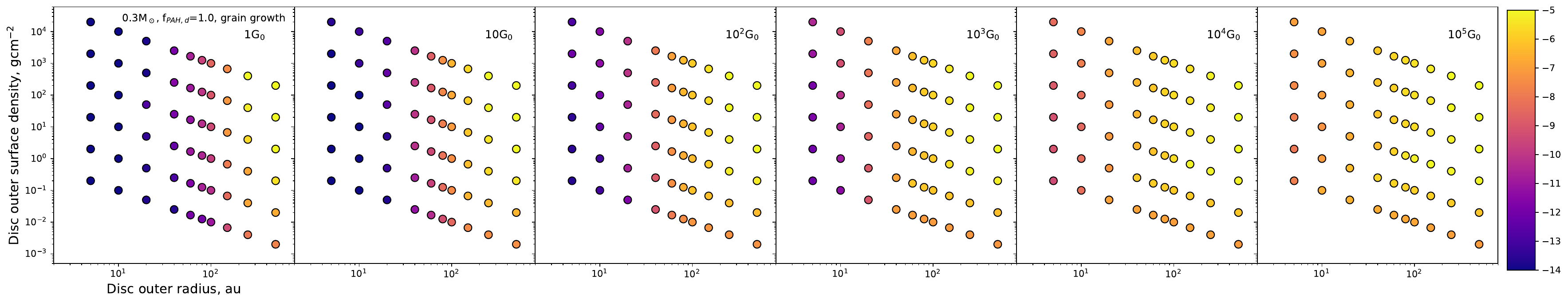}
    \includegraphics[width=2.0\columnwidth]{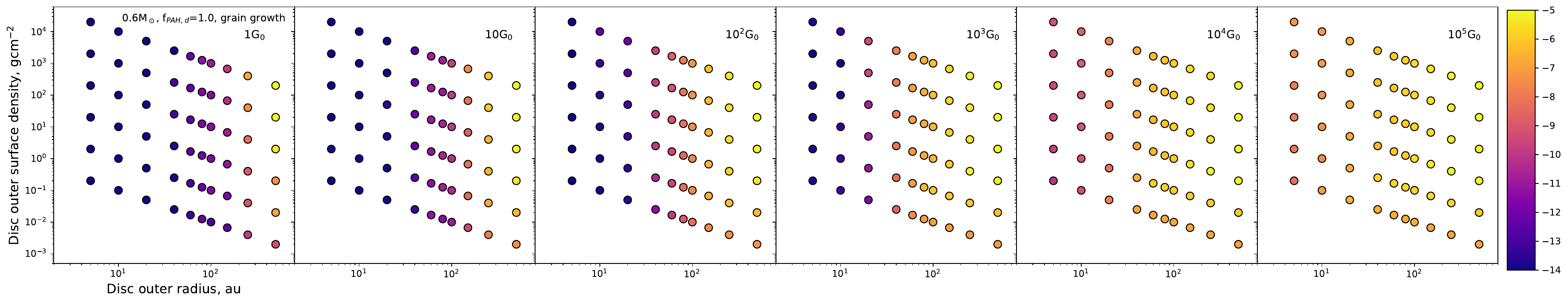}
    \includegraphics[width=2.0\columnwidth]{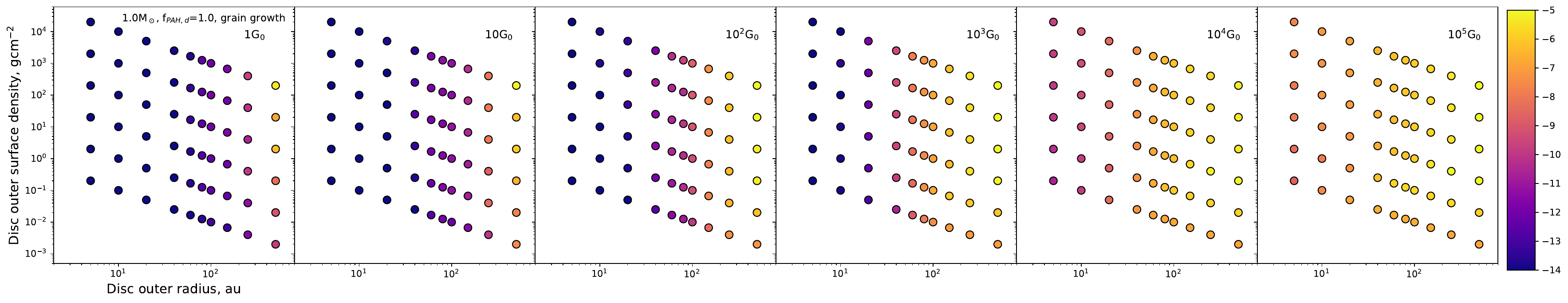}
    \includegraphics[width=2.0\columnwidth]{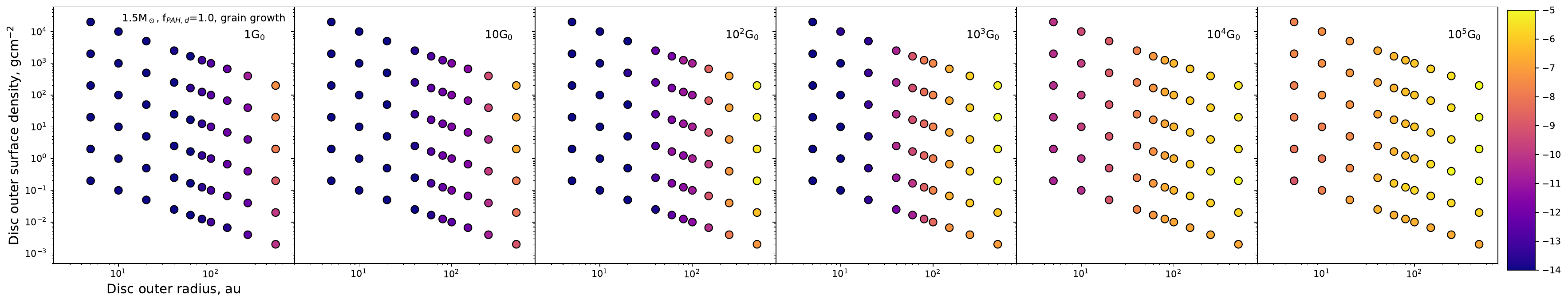}   
    \includegraphics[width=2.0\columnwidth]{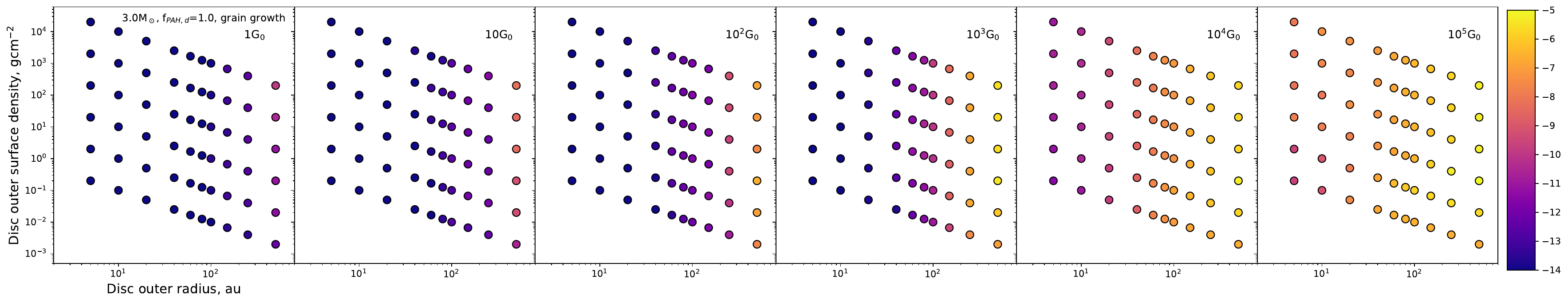}        
    \caption{The \textsc{fried} sub-grid that has $f_{\textrm{PAH,d}}=1$ and grain growth in the outer disc, depleting the wind of small grains. Each panel shows the mass loss rate ($\log_{10}[\dot{M}]$, M$_\odot$\,yr$^{-1}$, colour) as a function of disc outer radius and outer surface denisty. The stellar masses are 0.1, 0.3, 0.6, 1.0, 1.5 and 3.0\,M$_\odot$ from top to bottom and the UV radiation field is 1, 10, $10^2$, $10^3$, $10^4$, $10^5$G$_0$ from left to right. }
    \label{fig:PAH1p0grow}
\end{figure*}

Here in Figures \ref{fig:PAH1p0grow}--\ref{fig:PAH0p1small} we include plots of the mass loss rate as a function of disc outer radius and surface density. 

\begin{figure*}
    \includegraphics[width=2.0\columnwidth]{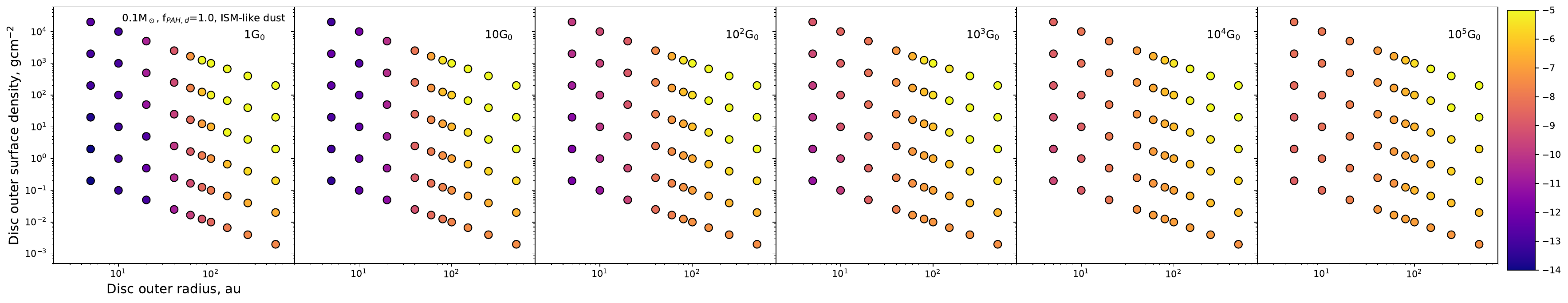}
    \includegraphics[width=2.0\columnwidth]{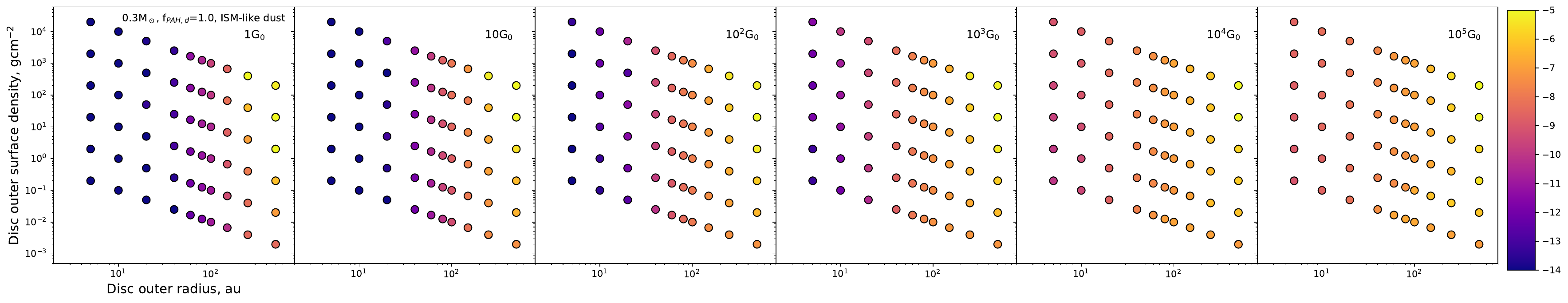}
    \includegraphics[width=2.0\columnwidth]{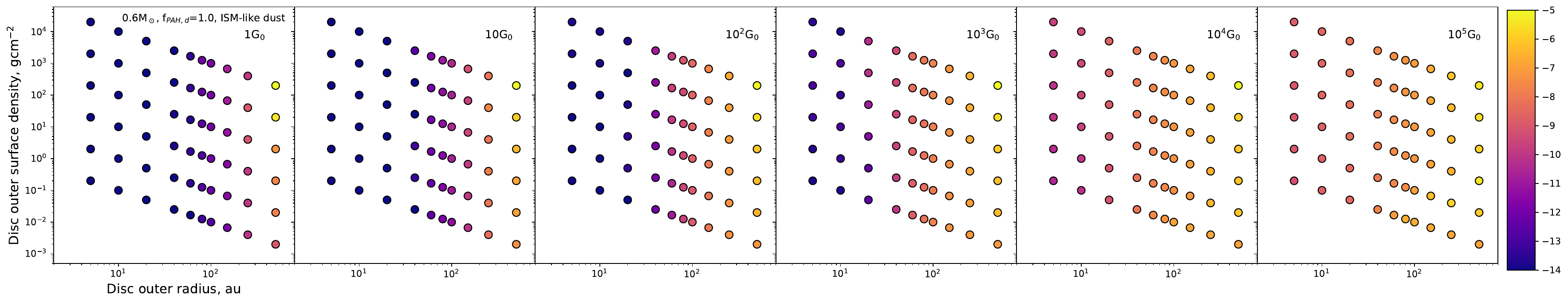}
    \includegraphics[width=2.0\columnwidth]{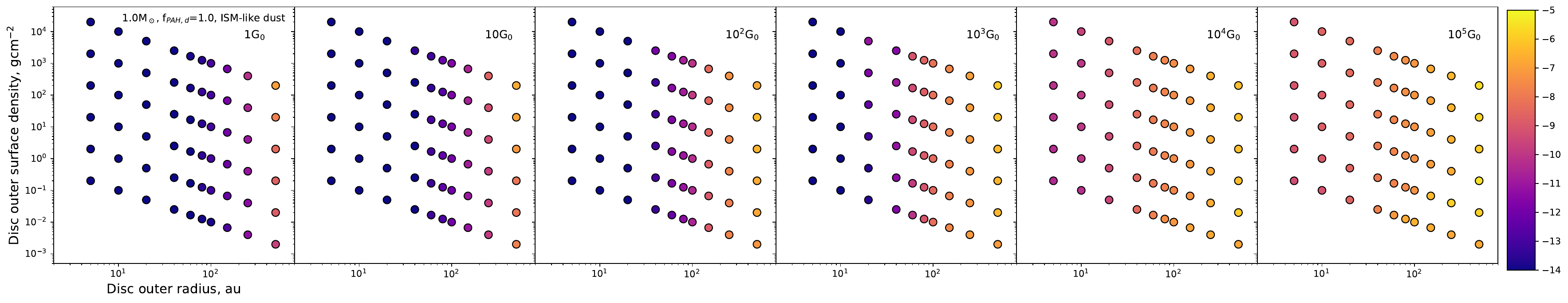}
    \includegraphics[width=2.0\columnwidth]{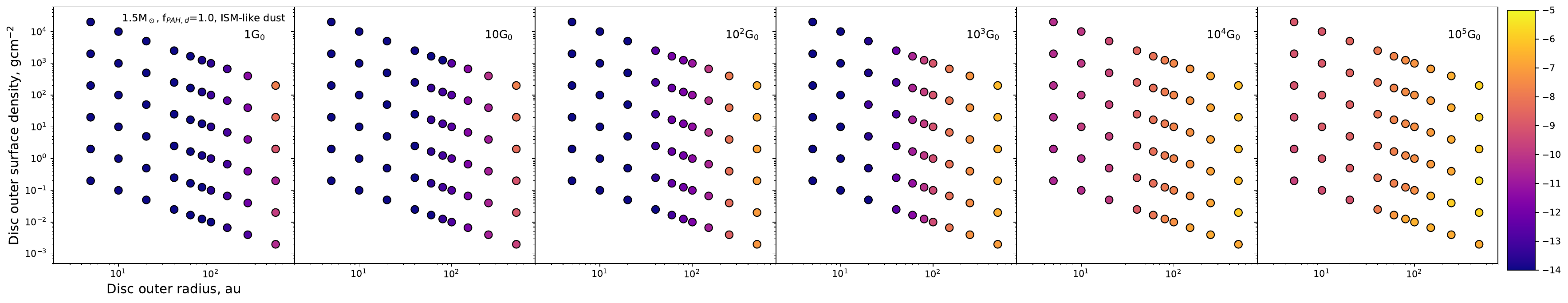}   
    \includegraphics[width=2.0\columnwidth]{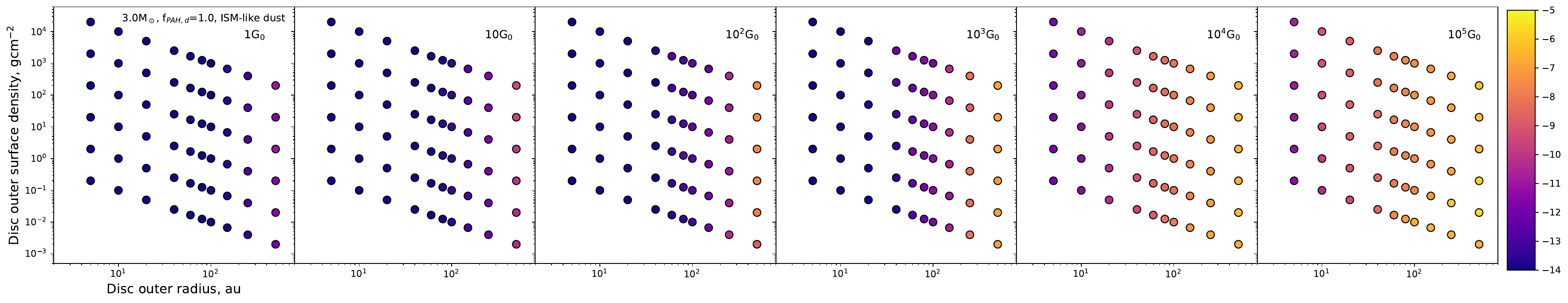}  
    \caption{The \textsc{fried} sub-grid that has $f_{\textrm{PAH,d}}=1.0$ and ISM-like dust in the outer disc. Each panel shows the mass loss rate ($\log_{10}[\dot{M}]$, M$_\odot$\,yr$^{-1}$, colour) as a function of disc outer radius and outer surface denisty. The stellar masses are 0.1, 0.3, 0.6, 1.0, 1.5 and 3.0\,M$_\odot$ from top to bottom and the UV radiation field is 1, 10, $10^2$, $10^3$, $10^4$, $10^5$G$_0$ from left to right.  }
    \label{fig:PAH1p0small}       
\end{figure*}

\begin{figure*}
    \includegraphics[width=2.0\columnwidth]{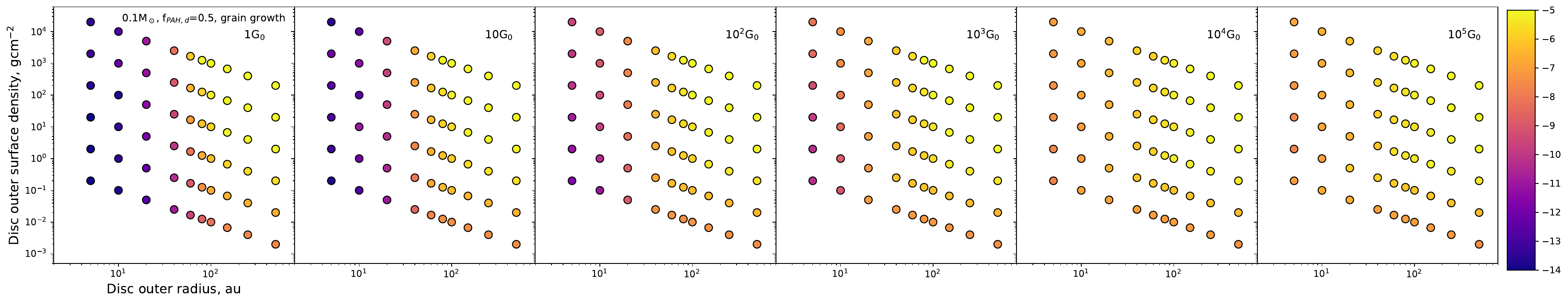}
    \includegraphics[width=2.0\columnwidth]{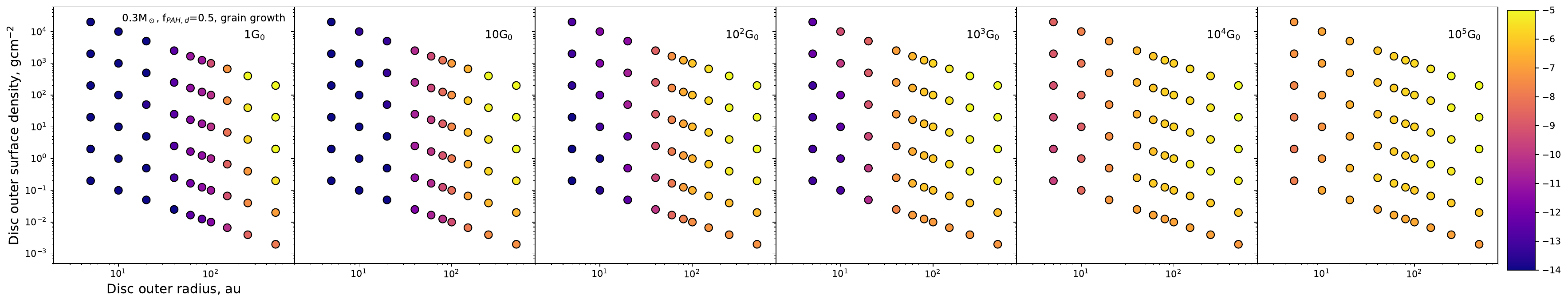}
    \includegraphics[width=2.0\columnwidth]{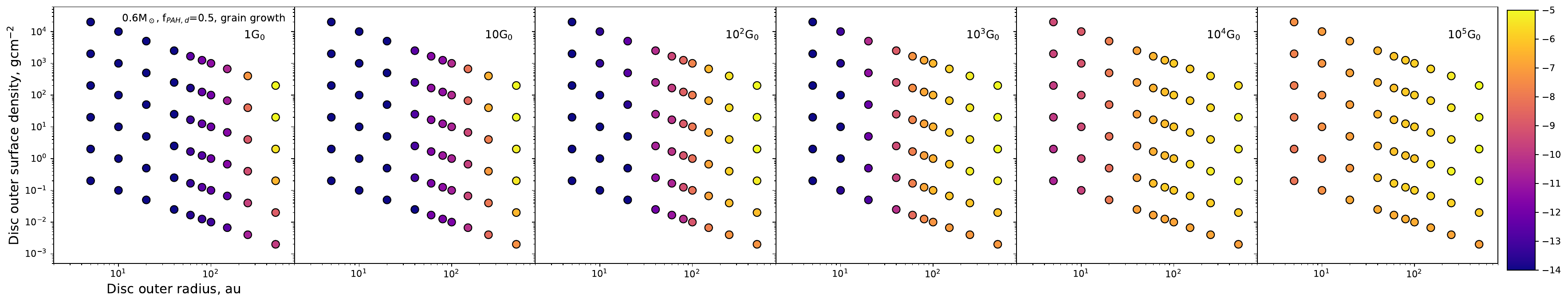}
    \includegraphics[width=2.0\columnwidth]{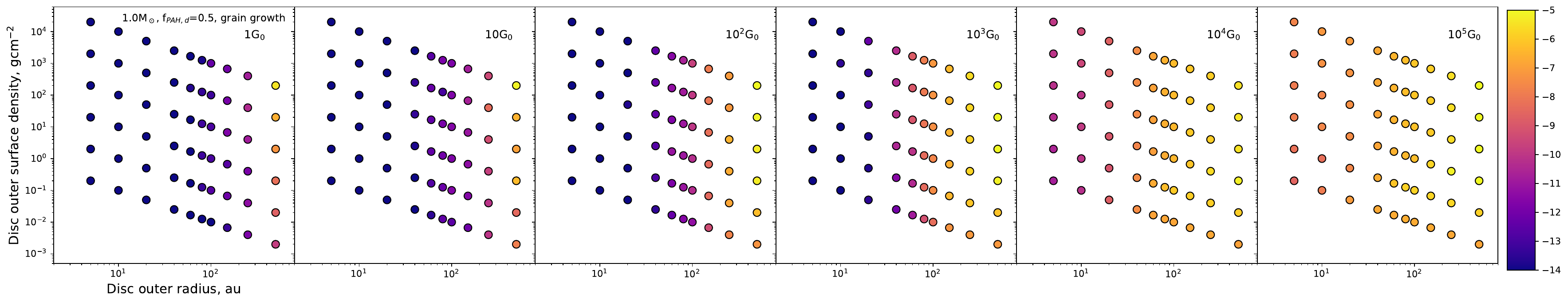}
    \includegraphics[width=2.0\columnwidth]{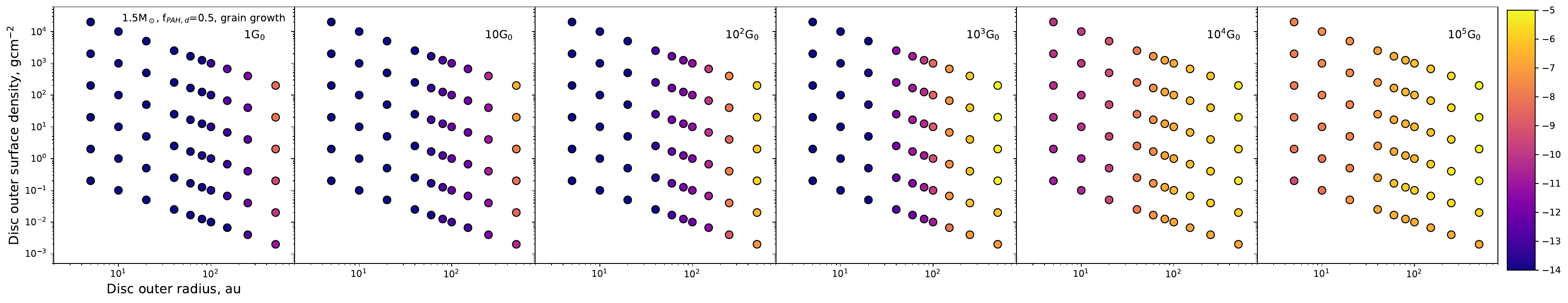}   
    \includegraphics[width=2.0\columnwidth]{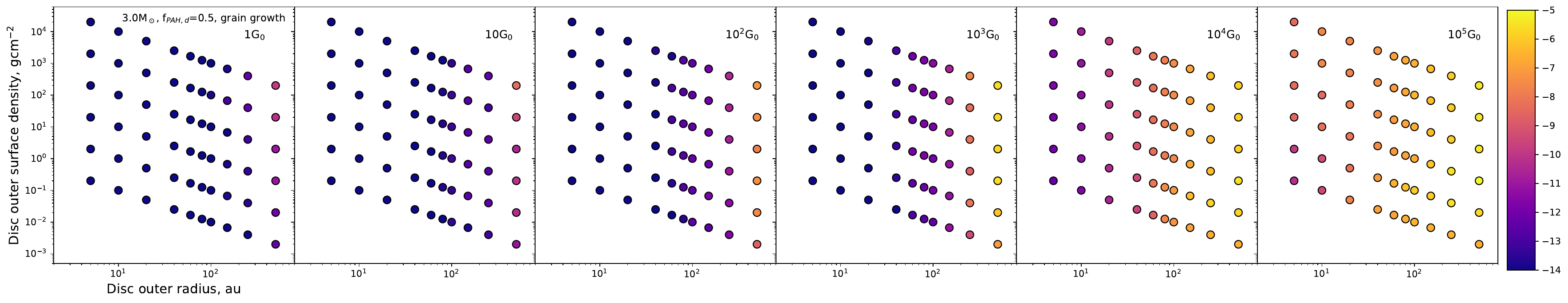}  
    \caption{The \textsc{fried} sub-grid that has $f_{\textrm{PAH,d}}=0.5$ and grain growth in the outer disc, depleting the wind of small grains. Each panel shows the mass loss rate ($\log_{10}[\dot{M}]$, M$_\odot$\,yr$^{-1}$, colour) as a function of disc outer radius and outer surface denisty. The stellar masses are 0.1, 0.3, 0.6, 1.0, 1.5 and 3.0\,M$_\odot$ from top to bottom and the UV radiation field is 1, 10, $10^2$, $10^3$, $10^4$, $10^5$G$_0$ from left to right. }
    \label{fig:PAH0p5grow}    
\end{figure*}

\begin{figure*}
    \includegraphics[width=2.0\columnwidth]{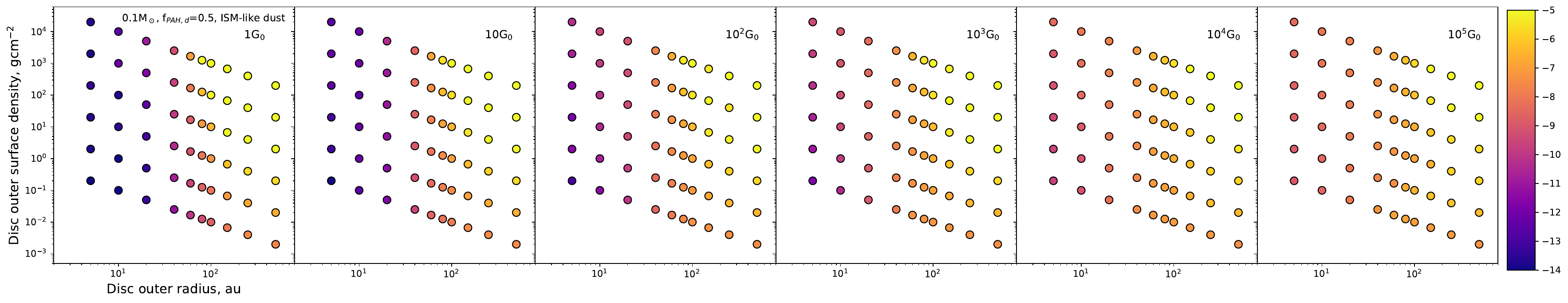}
    \includegraphics[width=2.0\columnwidth]{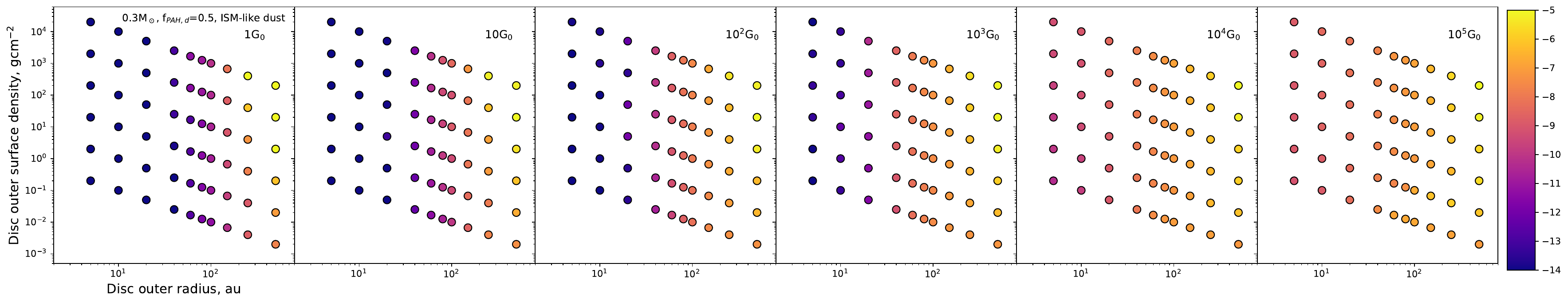}
    \includegraphics[width=2.0\columnwidth]{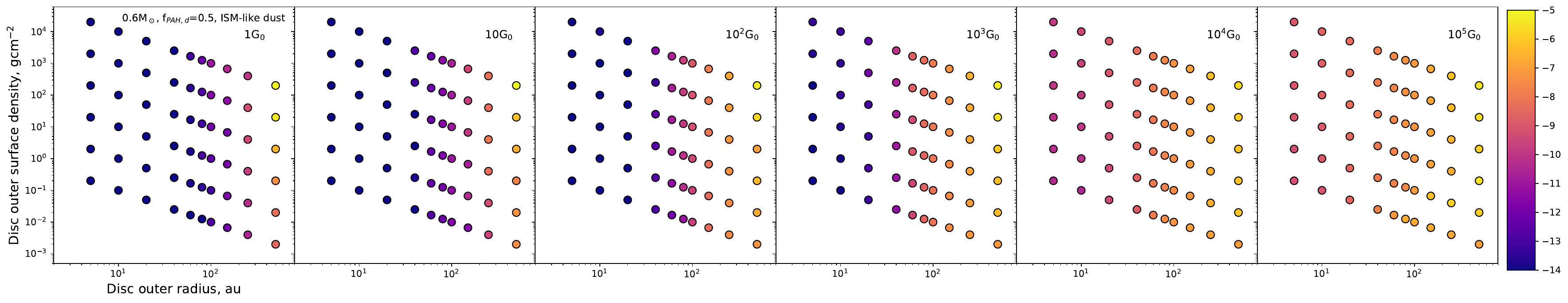}
    \includegraphics[width=2.0\columnwidth]{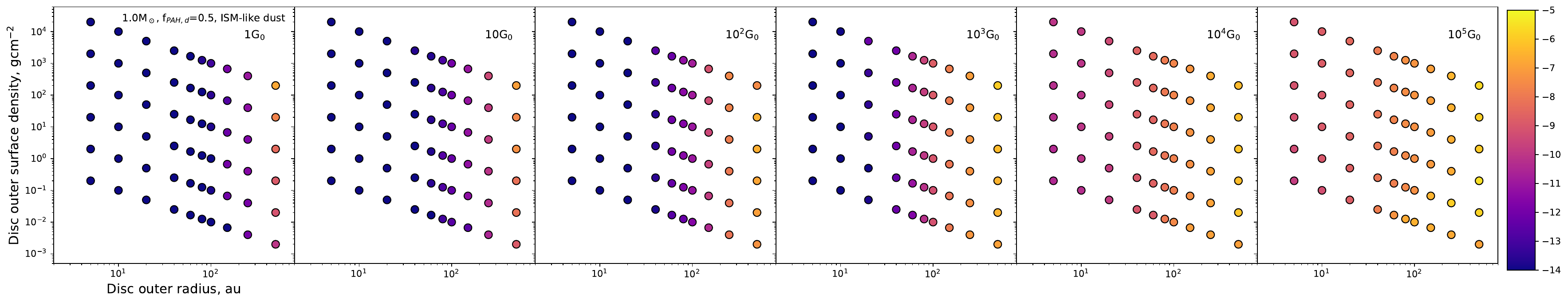}
    \includegraphics[width=2.0\columnwidth]{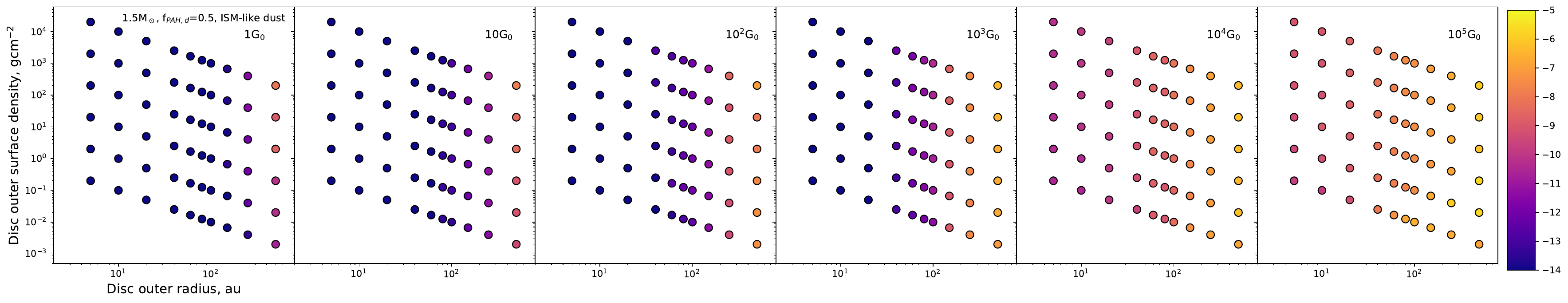}   
    \includegraphics[width=2.0\columnwidth]{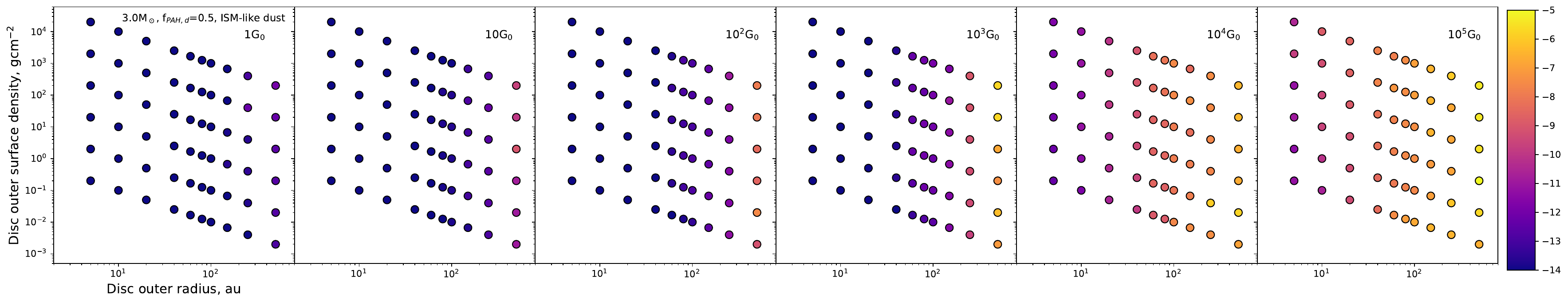}  
    \caption{The \textsc{fried} sub-grid that has $f_{\textrm{PAH,d}}=0.5$ and ISM-like dust in the outer disc. Each panel shows the mass loss rate ($\log_{10}[\dot{M}]$, M$_\odot$\,yr$^{-1}$, colour) as a function of disc outer radius and outer surface denisty. The stellar masses are 0.1, 0.3, 0.6, 1.0, 1.5 and 3.0\,M$_\odot$ from top to bottom and the UV radiation field is 1, 10, $10^2$, $10^3$, $10^4$, $10^5$G$_0$ from left to right.  }
    \label{fig:PAH1p0small}       
\end{figure*}

\begin{figure*}
    \includegraphics[width=2.0\columnwidth]{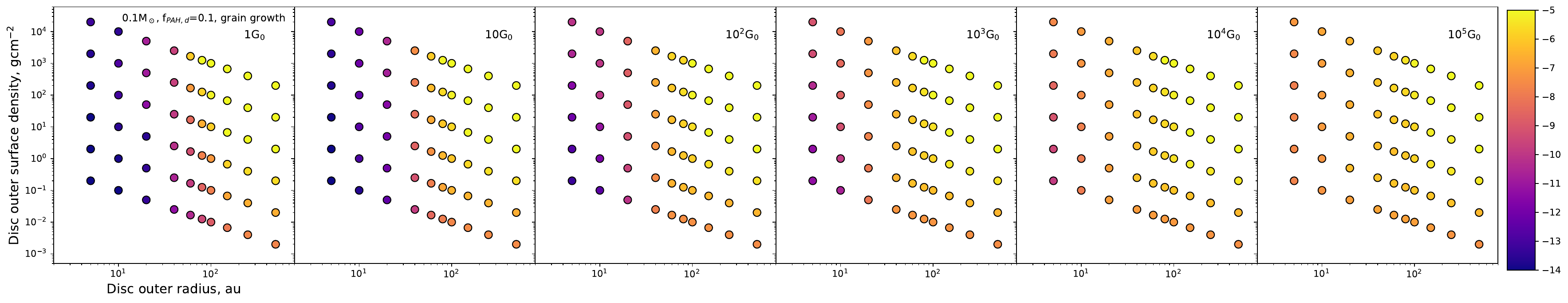}
    \includegraphics[width=2.0\columnwidth]{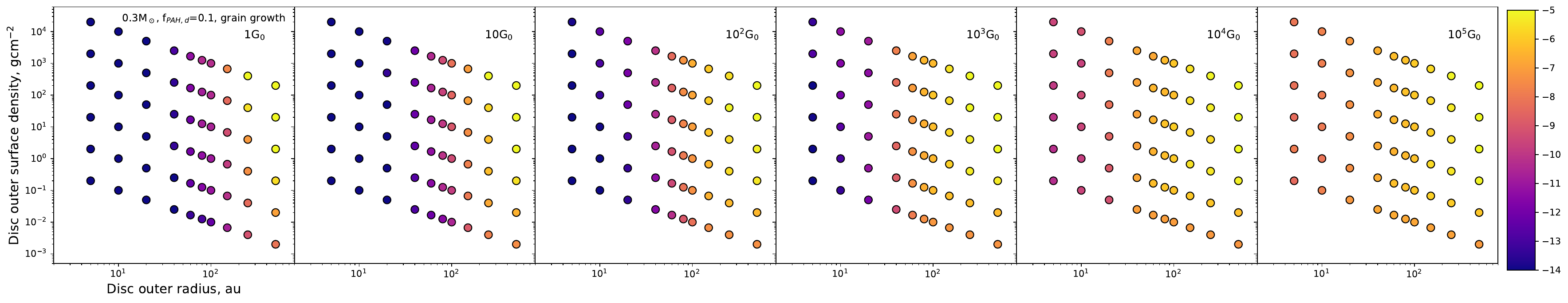}
    \includegraphics[width=2.0\columnwidth]{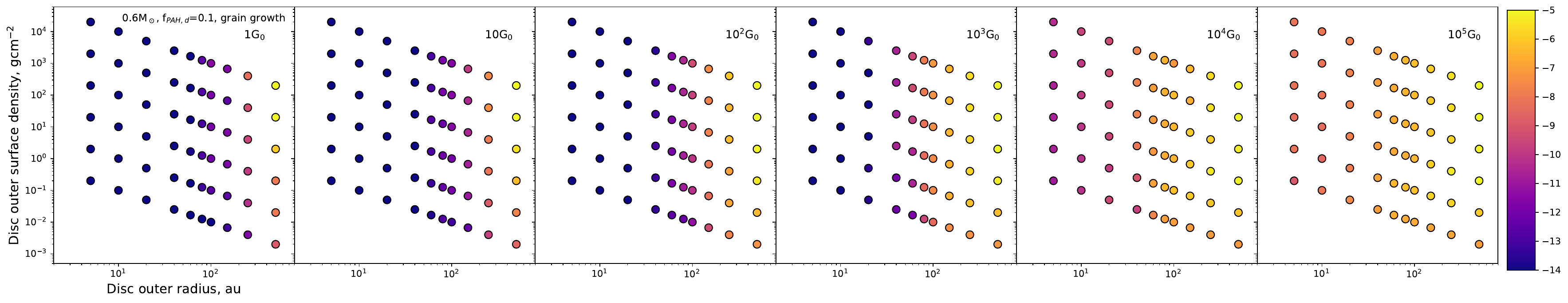}
    \includegraphics[width=2.0\columnwidth]{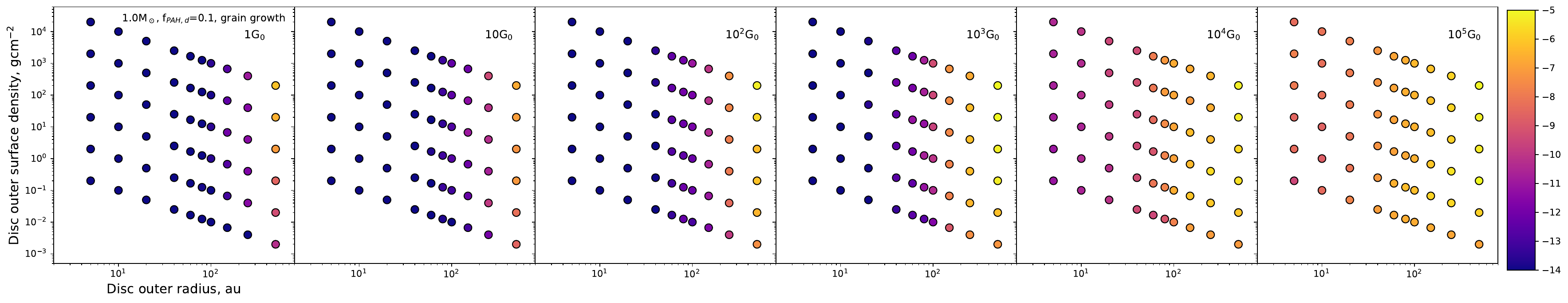}
    \includegraphics[width=2.0\columnwidth]{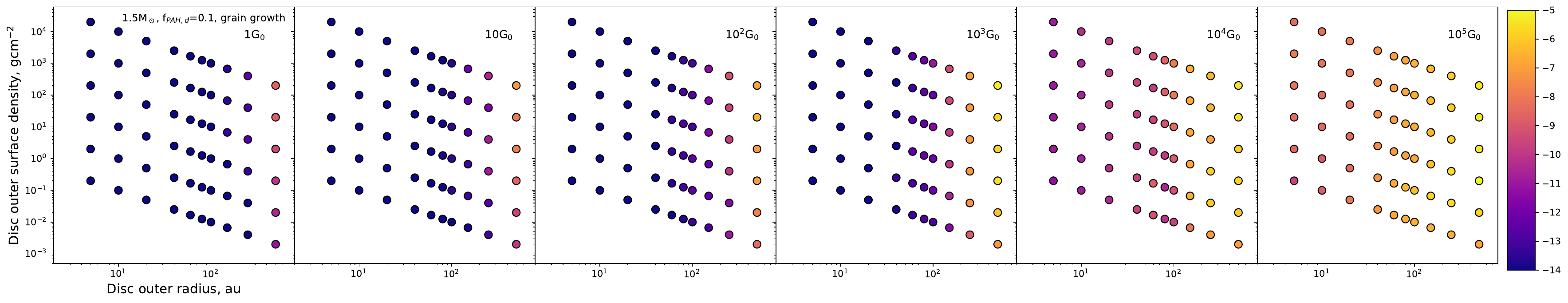}   
    \includegraphics[width=2.0\columnwidth]{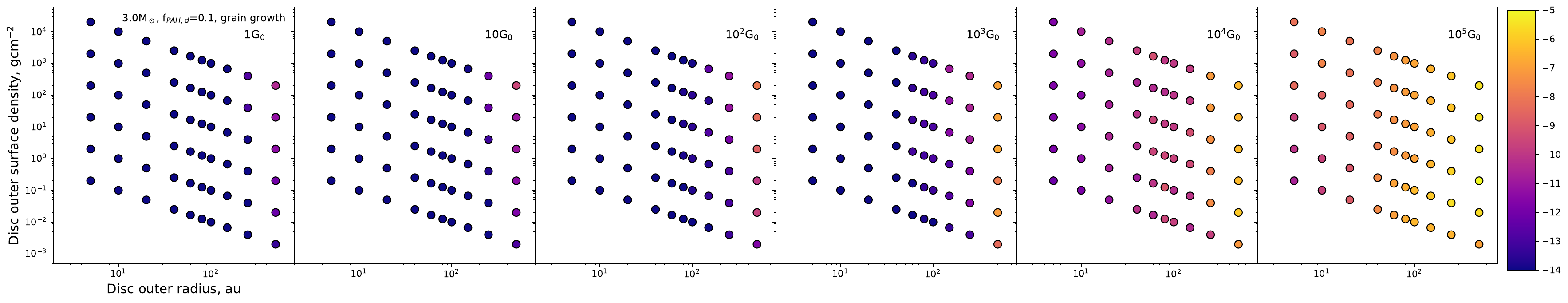}  
    \caption{The \textsc{fried} sub-grid that has $f_{\textrm{PAH,d}}=0.1$ and grain growth in the outer disc, depleting the wind of small grains. Each panel shows the mass loss rate ($\log_{10}[\dot{M}]$, M$_\odot$\,yr$^{-1}$, colour) as a function of disc outer radius and outer surface denisty. The stellar masses are 0.1, 0.3, 0.6, 1.0, 1.5 and 3.0\,M$_\odot$ from top to bottom and the UV radiation field is 1, 10, $10^2$, $10^3$, $10^4$, $10^5$G$_0$ from left to right. }
    \label{fig:PAH0p1grow}       
\end{figure*}

\begin{figure*}
    \includegraphics[width=2.0\columnwidth]{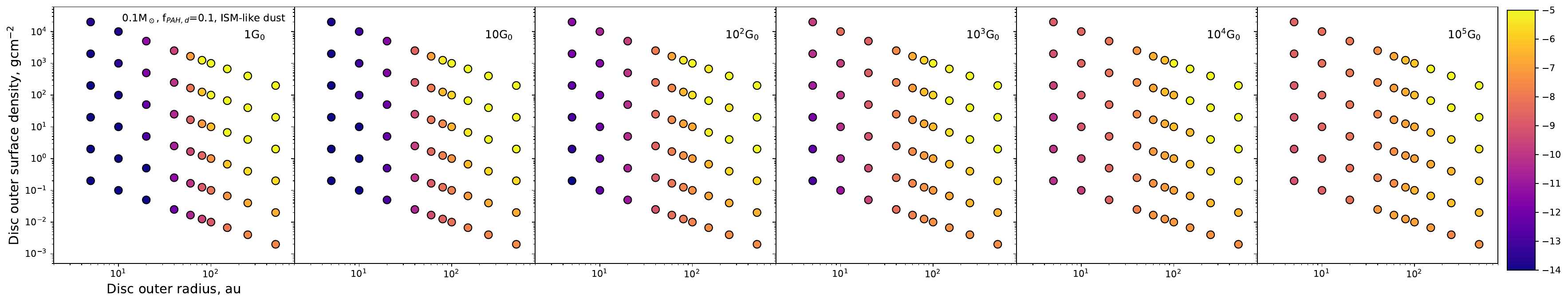}
    \includegraphics[width=2.0\columnwidth]{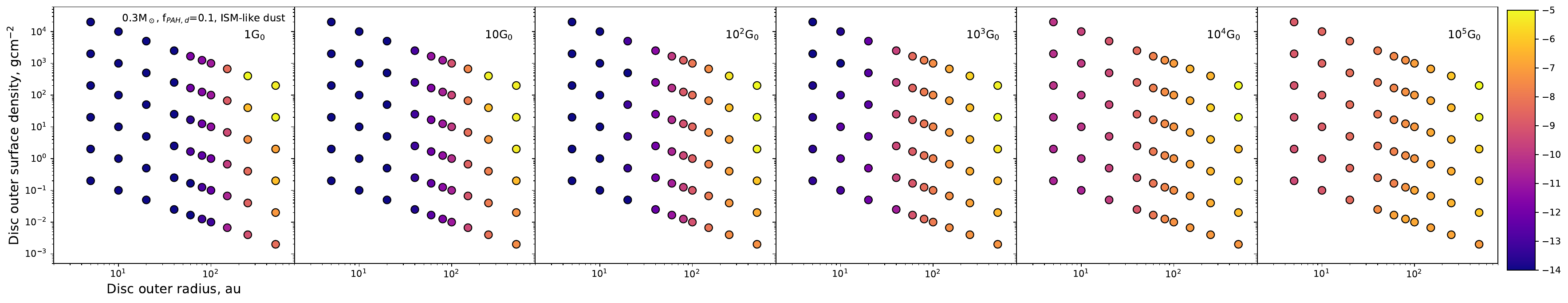}
    \includegraphics[width=2.0\columnwidth]{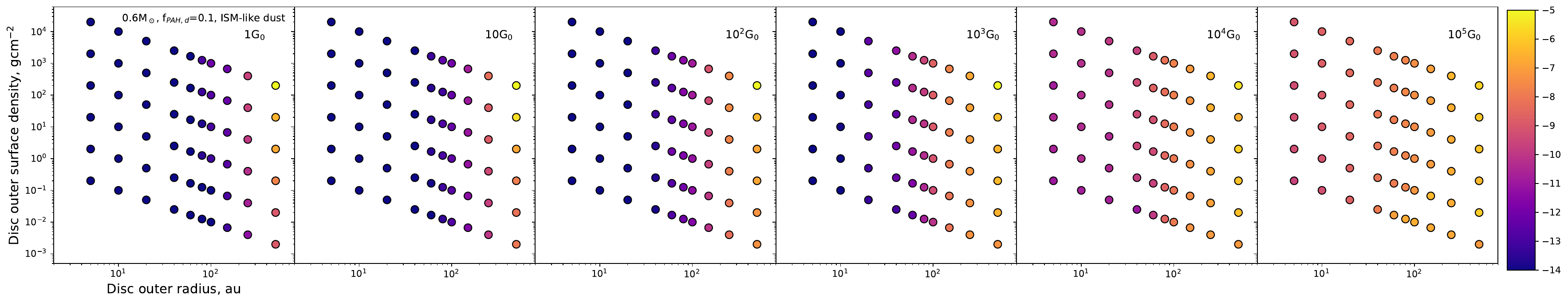}
    \includegraphics[width=2.0\columnwidth]{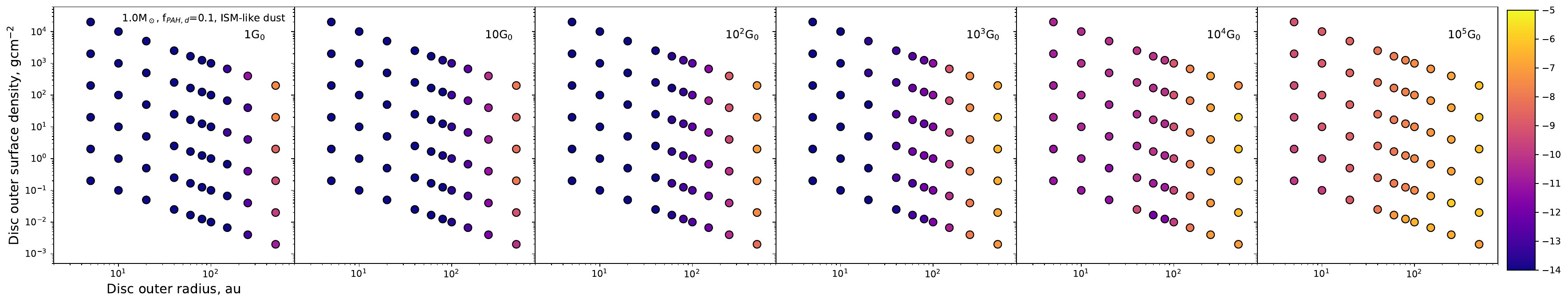}
    \includegraphics[width=2.0\columnwidth]{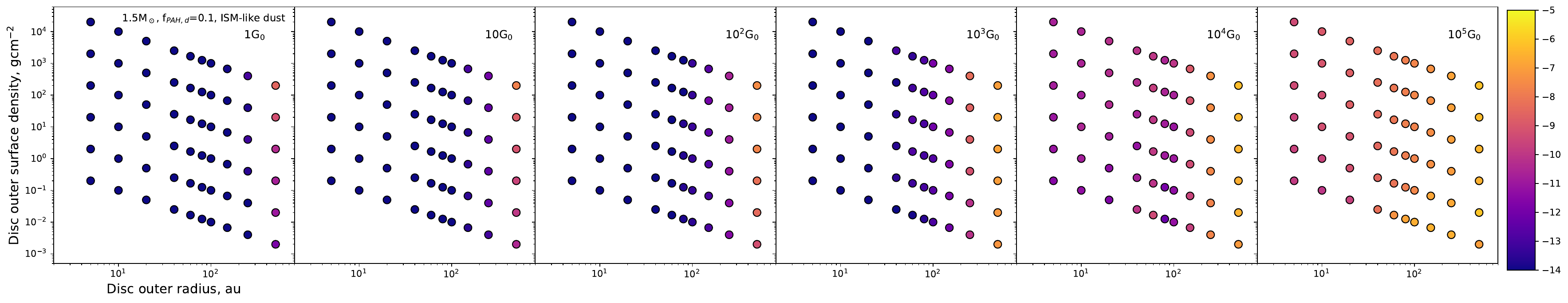}   
    \includegraphics[width=2.0\columnwidth]{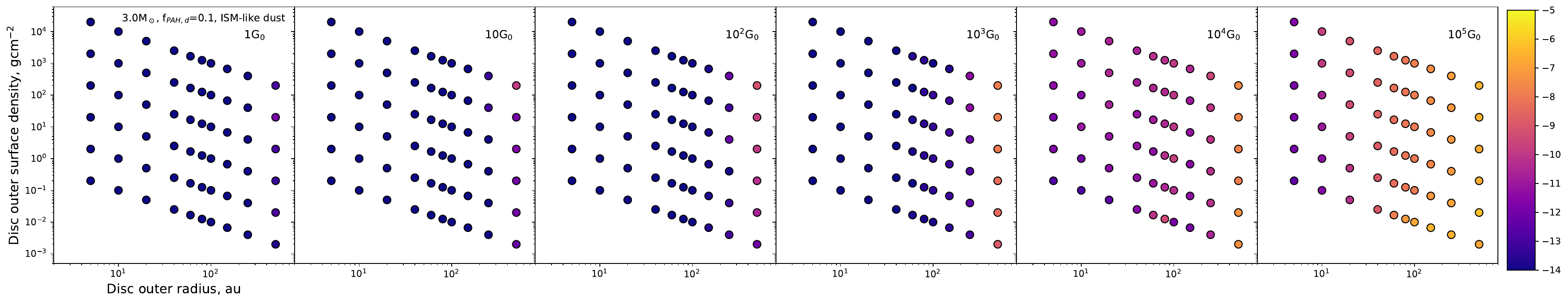}  
    \caption{The \textsc{fried} sub-grid that has $f_{\textrm{PAH,d}}=0.1$ and ISM-like dust in the outer disc. Each panel shows the mass loss rate ($\log_{10}[\dot{M}]$, M$_\odot$\,yr$^{-1}$, colour) as a function of disc outer radius and outer surface denisty. The stellar masses are 0.1, 0.3, 0.6, 1.0, 1.5 and 3.0\,M$_\odot$ from top to bottom and the UV radiation field is 1, 10, $10^2$, $10^3$, $10^4$, $10^5$G$_0$ from left to right. }
    \label{fig:PAH0p1small}       
\end{figure*}

\bsp	
\label{lastpage}
\end{document}